\documentstyle[12pt]{article}
\input {epsf}
\newcommand{\bfig}{\begin{figure}}
\newcommand{\efig}{\end{figure}}

\def\one-loop{\mbox{\scriptsize one-loop}}
\def\cD{{\cal D}}
\def\a{\alpha}

\def\l{\lambda}
\def\lb{\bar{\lambda}}
\def\s{\sigma}
\def\th{\theta}
\def\thb{\bar{\theta}}
\def\psib{\bar{\psi}}
\def\Db{\bar{D}}
\def\dg{\dagger}
\def\half{\frac{1}{2}}
\def\ud{\underline}
\def\bfpi{\mbox{\boldmath $\pi$}}
\def\G{\Gamma}

\def\e{\epsilon}
\def\eb{\bar{\epsilon}}

\catcode`\@=11

\@addtoreset{equation}{section}
\def\theequation{\arabic{section}.\arabic{equation}}

\def\@normalsize{\@setsize\normalsize{15pt}\xiipt\@xiipt
\abovedisplayskip 14pt plus3pt minus3pt%
\belowdisplayskip \abovedisplayskip
\abovedisplayshortskip  \z@ plus3pt%
\belowdisplayshortskip  7pt plus3.5pt minus0pt}

\def\small{\@setsize\small{13.6pt}\xipt\@xipt
\abovedisplayskip 13pt plus3pt minus3pt%
\belowdisplayskip \abovedisplayskip
\abovedisplayshortskip  \z@ plus3pt%
\belowdisplayshortskip  7pt plus3.5pt minus0pt
\def\@listi{\parsep 4.5pt plus 2pt minus 1pt
            \itemsep \parsep
            \topsep 9pt plus 3pt minus 3pt}}

\def\underline#1{\relax\ifmmode\@@underline#1\else
        $\@@underline{\hbox{#1}}$\relax\fi}
\@twosidetrue

\relax

\catcode`@=12

\catcode`\@=11

\def\section{\@startsection{section}{1}{\z@}{3.5ex plus 1ex minus
   .2ex}{2.3ex plus .2ex}{\large\bf}}

\def\thesection{\Roman{section}.}

\def\appendix{\setcounter{section}{0}
        \def\thesection{Appendix }
        \def\theequation{\Alph{section}.\arabic{equation}}}

\def\ps@headings{\def\@oddfoot{}\def\@evenfoot{}
\def\@oddhead{\hbox{}\hfill
        \makebox[.5\textwidth]{\raggedright\ignorespaces --\thepage{}--
        \hfill {}}}
\def\@oddhead{\hbox{}\hfill --\thepage{}-- \hfill
        {}}
\def\@evenhead{\@oddhead}
\def\subsectionmark##1{\markboth{##1}{}}
}

\ps@headings

\catcode`\@=12

\relax

\def\figcap{\section*{Figure Captions\markboth
        {FIGURECAPTIONS}{FIGURECAPTIONS}}\list
        {Fig. \arabic{enumi}:\hfill}{\settowidth\labelwidth{Fig. 999:}
        \leftmargin\labelwidth
        \advance\leftmargin\labelsep\usecounter{enumi}}}
 \relax
\def\tablecap{\section*{Table Captions\markboth
        {TABLECAPTIONS}{TABLECAPTIONS}}\list
        {Table \arabic{enumi}:\hfill}{\settowidth\labelwidth{Table 999:}
        \leftmargin\labelwidth
        \advance\leftmargin\labelsep\usecounter{enumi}}}
 \relax
\def\reflist{\section*{References\markboth
        {REFLIST}{REFLIST}}\list
        {[\arabic{enumi}]\hfill}{\settowidth\labelwidth{[999]}
        \leftmargin\labelwidth
        \advance\leftmargin\labelsep\usecounter{enumi}}}
 \relax

\catcode`\@=11

\def\ps@headings{\def\@oddfoot{}\def\@evenfoot{}
\def\@oddhead{\hbox{}\hfill
        \makebox[.5\textwidth]{\raggedright\ignorespaces --\thepage{}--
        \hfill {}}}
\def\@evenhead{\@oddhead}
\def\subsectionmark##1{\markboth{##1}{}}
}

\ps@headings

\relax

\newskip\humongous \humongous=0pt plus 1000pt minus 1000pt

\newif\ifdtup

\def\beq{\begin{equation}}
\def\eeq{\end{equation}}

\def\beqn{\begin{eqnarray}}
\def\eeqn{\end{eqnarray}}
\relax

\def\G2{{\; \rm GeV/}c^2}
\def\G{\; \rm GeV}

\def\dotx{\dotx{\dot\overline{x}}}

\relax

\def\p{\partial}

\begin{document}
\begin{titlepage}
\begin{flushright}
       {\normalsize OU-HET 287 \\  hep-th/9801084\\
         January, 1998 }
\end{flushright}
%
\begin{center}
  {\large \bf $USp(2k)$ Matrix Model:  \\
   Nonperturbative Approach to Orientifolds  }
\footnote{This work is supported in part
 by the Grant-in-Aid  for Scientific Research Fund (2126,97319)
from the Ministry of Education, Science and Culture, Japan.}

\vfill
         {\bf H.~Itoyama}  \\
            and \\
              {\bf A.~Tokura}\\
        Department of Physics,\\
        Graduate School of Science, Osaka University,\\
        Toyonaka, Osaka, 560 Japan\\
\end{center}
\vfill
\begin{abstract}
We discuss theoretical implications of the large $k$ $USp(2k)$
matrix model in zero dimension. The model appears as the matrix model
of type $IIB$ superstrings
 on a large $T^{6}/{\cal Z}^{2}$ orientifold via the
matrix twist operation. In the small volume limit, the model behaves
four dimensional  and its $T$ dual is six-dimensional worldvolume
  theory of type $I$ superstrings in ten spacetime dimensions.
Several theoretical considerations including the analysis on planar diagrams,
the commutativity of the projectors with supersymmetries and the cancellation
of gauge anomalies are given, providing us with the rationales for the
choice of the Lie algebra and the field content.
A few classical solutions are constructed which correspond to Dirichlet
 $p$-branes and some fluctuations are evaluated.
The particular scaling limit with matrix $T$ duality transformation
is discussed which derives
 the $F$ theory compactification on an elliptic fibered $K3$.
\end{abstract}
\vfill
\end{titlepage}

\section{Introduction}

 This paper discusses in depth theoretical implications of the
 $USp(2k)$ matrix model in zero dimension introduced in ref.~\cite{IT}. 
 A particular emphasis will be given to the aspects of the model
 as a nonperturbative framework to deal with orientifold 
 compactification.

 Gauge fields and strings have governed our thoughts on unified theory of all
 forces including gravity and constituents for more than two decades. One of
 our current theoretical endeavours is, it seems, to take gauge
 fields as dynamical variables of noncommuting matrix coordinates\cite{Witten}
 to construct string theory from matrices. 
 This approach strives for overcoming some of the difficulties of the first
 quantized superstring theory, which 
 have led to an inevitable impasse: one may list, among other things,
 the existence of infinitely degenerate perturbative vacua, the problem of
 supermoduli {\it etc}.  One dimensional matrix model \cite{BFSS}
 of $M$ theory \cite{M} has
 obtained  successes  on the agreement of the spectrum and other properties
 with the low energy eleven-dimensional
 supergravity theory while the zero-dimensional model \cite{KEK} of type
 $IIB$ superstrings lays its basis on the 
 correspondence \cite{Nambu,Bars,FFZ,KEK} with
 the first quantized action of the Schild type gauge \cite{Schild} and appears
 to be numerically accessible. We will often refer to the latter case as
 reduced model.
 See refs. \cite{subdev} for some of the references on the subsequent
 developments.

   We would like to show that the reduced model presented in this paper 
  descends from the first quantized  nonorientable
 type $I$ superstring  theory
 \cite{GSano},  which is believed to be related to heterotic string theory
 \cite{GHMR}  by S duality \cite{M,PW}.
In this sense,  it is expected that the model is exposed to phenomenological
 questions of particle physics by the presence of gauge bosons,
 matter fermions and other properties.
As is pointed out in ref. \cite{IT},
 the model, at the same time, captures  one of the
 exact results in string theory, namely
 the $F$ theory compactification on
 an elliptic fibered $K3$, which is originally deduced geometrically
 \cite{V} from the $SL(2,Z)$  duality \cite{SL(2Z)}:
 it is nonetheless exact quantum mechanically.
 
 In the next section, the definition of the $USp(2k)$ matrix model introduced
 in ref. \cite{IT} is recalled. The relationship of the parts not
 involving the fields in the fundamental representation with the type
  $IIB$ matrix model is given precisely, by introducing projectors onto
 the $USp$ adjoint as well as the antisymmetric representation. This is
 found to be useful in developing the analysis in  remaining sections.
 The definition of the model appears to be rather ad hoc at first.
 In the subsequent three sections, we will show that our model passes in fact
several stringent criteria which the large $k$ reduced model  of
 orientifold must satisfy.
 We will be able to provide the rationales for our choice of the
 Lie algebra $usp$, for the choice of the number of the noncommuting
 coordinates belonging to the adjoint representation and that to the
 antisymmetric representation,
 and finally for the number of multiplets needed, denoted by $n_{f}$,
 belonging to the fundamental representation.

The most basic notion of the large $k$  reduced models
 is that the dense set of Feynman diagrams in the large $k$ limit
 forms the string worldsheet \cite{tH}.
This is not limited to a combinatorial equivalence.
The reduced $U(2k)$ Yang-Mills model goes to the string action
in the Schild gauge.
The Lie algebra $u(2k)$ becomes isomorphic to the area preserving
diffeomorphisms on a sphere.
In section 3, we begin with showing how this fact is extended
to nonorientable strings.
We examine the role played by the matrix $F$
in large $k$ $USp$ Feynman diagrams, ignoring the diagrams coming
  from the fields belonging to the fundamental representation.
 This is combined with the analysis relating $F$ to the worldsheet
 involution in the large $k$ limit, telling us that the surfaces created by
  the dense set of Feynman diagrams are nonorientable.
  The correspondence with the first quantized operator approach confirms that
 $F$ is a matrix analog of the twist operation.
   This is strengthened by showing that  equation (\ref{eq:rel})
 changes sign under the matrix T duality
 transformation \cite{WT}.
   In section $4$, we examine the commutativity of the projectors with
dynamical as well as kinematical supersymmetry.  The cases  which pass this
criterion with eight dynamical and eight kinematical supercharges are found to
be very scarce. The field content of our model stands as the most natural
choice. In section $5$, we discuss the role played by the fields in the
 fundamental representation  and the cancellation of gauge anomalies.
Obviously, these fields create boundaries of the surfaces.

Combining these analyses in sections $3$, $4$ and $5$,
 we conclude that the model in
 its original form is the large $k$ reduced model of type $IIB$ superstrings
 on a large $T^{6}/{\cal Z}^{2}$
 orientifold.  In the other limit, namely the small volume limit in which
   the model behaves as in four dimensional flat spacetime, the $T$ duality
 transformation takes this model into the six-dimensional worldvolume theory
  representing type $I$ superstrings in ten spacetime dimensions.
  The anomaly cancellation of this
 worldvolume gauge theory in section $5$ selects $n_{f}=16$, telling us
 that this is the matrix counterpart of the original Green-Schwarz
 cancellation leading to $SO(32)$ type $I$ nonorientable superstrings.
 
 In section $6$, we turn to constructing classical solutions which
 correspond to a D-string  and two (anti-)parallel D-strings.
 A formula for the one-loop effective action 
 on a general background is obtained.
 This is used to evaluate the potential between two antiparallel D-strings.
  Evidently,  two additional dimensions are not generated in this naive
 large $k$ limit. 
 These solutions are straightforwardly generalized to solutions
 representing a D$p$-brane and parallel D$p$-branes, which we illustrate
 in the case of $p=3$ in section $7$.
  In section $8$, applying some of the results obtained 
 in sections $3$, $4$ and $5$,
  we supplement the discussion of ref. \cite{IT} on the connection with
the  $F$ theory compactification on
 an elliptic fibered $K3$.

\section{Definition of the $USp(2k)$ matrix model}

 We adopt a notation   that  the inner product of the two $2k$ dimensional
 vectors  $u_{i}$ and $v_{i}$ invariant under $USp(2k)$  are
\beqn
 \langle u, v \rangle = u_{i} F^{ij} v_{j} \;\;, \;\; \\
  F^{ij} =  \left(
                  \begin{array}{cc}
                  0 & I_k \\
                  -I_k & 0
                  \end{array}
            \right)
\eeqn
$I_k$ is the unit matrix.
 The raising and lowering of the indices are  done  by $F=F^{ij}$
  and $F^{-1} = F_{ij}$.
   The element $X$ of the $usp(2k)$ Lie algebra
  satisfying  $X^{t}F + FX =0$ and $X^{\dagger} = X$ can be represented as
\beqn
  X =\left(
     \begin{array}{cc}
           M     &     N    \\
           N^{*} &  -M^{t}
     \end{array}
            \right)
\eeqn
  with $M^{\dagger} =M$ and $N^{t} =N $.
 It is sometimes convenient to adopt the tensor product notation:
\beq
X = \left( \frac{1+ \s^3}{2} \right) \otimes M
 + \left( \frac{1- \s^3}{2} \right) \otimes (-M^t)
 + \s^{+} \otimes N
 + \s^{-} \otimes N^{*}
\label{eq:adjoint matrix}
\eeq
where $\s^1$, $\s^2$ and $\s^3$
are Pauli matrices, and $\s^{\pm} \equiv (\s^1 \pm i \s^2)/2$.
On the other hand, the element $Y$ of the antisymmetric representation
of the $USp(2k)$ is
\beq
 Y  = \left(
         \begin{array}{cc}
         A  &  B \\
         C  &  A^{t}
        \end{array}
    \right)
\label{eq:asym matrix}
\eeq
with $B^t = - B$ and $C^t= -C$.
The hermiticity condition can be imposed.
In the tensor product notation, eq. (\ref{eq:asym matrix}) becomes then
\beq
 \left( \frac{1+ \s^3}{2} \right) \otimes A
 + \left( \frac{1- \s^3}{2} \right) \otimes A^t
 + \s^{+} \otimes B
 + \s^{-} \otimes (-B^{*})
\label{eq:asym hermite matrix}
\eeq
with $A^{\dagger}=A$ and $B^t = -B $.

  Let us recall the definition of  the  $USp(2k)$ matrix model
  in zero dimension introduced in ref. \cite{IT}.

    Our zero-dimensional model can be written, by borrowing
 ${\cal N}=1$, $d=4$ superfield notation in the Wess-Zumino gauge. One  simply drops
   all spacetime dependence of the fields but keeps all grassmann coordinates
  as they are:
\beqn
\label{eq:def}
   S &\equiv&  S_{ {\rm vec}} + S_{{\rm asym}} + S_{{\rm fund}} \;\;
\\
  S_{ {\rm vec}} &=& \frac{1}{4 g^{2}} \;  Tr \left(
 \int d^2 \th W^{\a} W_{\a} + h.c. +
4 \int d^2 \th d^2 \thb \Phi^{\dg} e^{2V} \Phi e^{-2V} \right)
 \nonumber   \\
  S_{{\rm asym}} &=&    \frac{1}{g^{2}} \int d\theta^{2} d \bar{\theta}^{2}
 \left( T^{* \; ij} \left( e^{2V({\rm asym} ) }
 \right)_{ij}^{\;\;\;k \ell}
  T_{k \ell} +  \tilde{T}^{ij} \left( e^{-2V({\rm asym} ) }
 \right)_{ij}^{\;\;\;k \ell}
 \tilde{T}^{* }_{k \ell} \right)
\nonumber\\
& & + \frac{1}{g^{2}} \left\{ \sqrt{2} \int d\theta^{2}
\tilde{T}^{ij} \left( \Phi_{(asym)}
 \right)_{ij}^{\;\;\;k \ell}
T_{k \ell}   + h.c.
 \right\}
\nonumber
\\
  S_{{\rm fund}} &=&  \frac{1}{g^{2}} \sum_{f=1}^{n_f}
\left[
  \int d^2 \th d^2 \thb
\left( Q_{(f)}^{* \;i} \left( e^{2V} \right)_i^{\;\;j} Q_{(f) \; j}
+  \tilde{Q}_{(f)}^{ i} \left( e^{-2V} \right)_i^{\;\;j}
        \tilde{Q}_{(f) \; j}^{*} \right)
\right.
\nonumber
\\
& &
\qquad
 +
\left.
\left\{
\int d^2 \th
\left(
        m_{(f)} \tilde{Q}_{(f)}^{\;\;\;\; i}  Q_{(f) \;i}
        + \sqrt{2} \tilde{Q}_{(f)}^{\;\;\;\; i}
                 \left( \Phi \right)_i^{\;\;j} Q_{(f) \;j}
\right)
+ h.c.
\right\}
\right]  \;\;\;.
   \nonumber
\eeqn
  The chiral superfields introduced above  are
\beqn
  W_{\alpha} &=&  -\frac{1}{8} \Db \Db e^{-2V} D_{\a} e^{2V}
\;,\;
\Phi = \Phi + \sqrt{2} \th \psi_{\Phi} + \th \th F_{\Phi}
 \;\;,\;\;
\\
Q_{i}&=& Q_{i} + \sqrt{2} \th \psi_{Q \; i} + \th \th F_{Q \; i}
\;, \;
   T_{ij} = T_{ij} + \sqrt{2} \th \psi_{T \; ij} + \th \th F_{T \; ij}
 \;\;,\;\; \\
{\rm while}\;\;
D_{\a} &=& \frac{\p}{\p \th^{\a}} \;\; , \;\;
\bar{D}_{\dot{\a}} =  -\frac{\p}{\p \thb^{\dot{\a}}}
\\
V &=& - \th \s^m \thb v_m + i \th \th \thb \lb - i \thb \thb \th
\l + \half \th \th \thb \thb D \;\;.
\eeqn
  We represent the antisymmetric tensor superfield $T_{ij}$
  as
\beqn
\label{eq:antisym}
 Y \equiv \left( TF \right)_{i}^{\;j}
 = \left(
         \begin{array}{cc}
         A  &  B \\
         C  &  A^{t}
        \end{array}
    \right)
\label{eq:Y}
\eeqn
  with $B^{t} = -B$, $C^{t}=- C$.  We define $\tilde{Y}$ similarly.

  In terms of components, the action reads, with indices suppressed,
\beqn
S_{vec}
&=&
\frac{1}{g^2} Tr (- \frac{1}{4} v_{m n} v^{m n} -
[\cD_{m}, \Phi]^{\dagger} [ \cD^{m}, \Phi ]
- i \lambda \s^{m}
[ \cD_{m} , \overline{\lambda} ]
- i \overline{\psi}  \overline{\s}^{m}
[\cD_{m} , \psi]
\nonumber \\
& &
\qquad
- i \sqrt{2} [ \l , \psi ] \Phi^{\dagger} - i \sqrt{2}
[ \overline{\lambda}   , \overline{\psi} ] \Phi)
\nonumber \\
& &
+ \frac{1}{g^2} Tr \left(
\frac{1}{2} D D  - D [\Phi^{\dagger} , \Phi] + F_{\Phi}^{\dagger} F_{\Phi}
\right)
\eeqn

\beqn
S_{asym} &=&
\frac{1}{g^2}
\left\{
- ( \cD_{m} T )^{*} (\cD^{m} T )
- i {\overline{\psi}}_{T}  \overline{\s}^{m}
\cD_{m} \psi_{T}
- i \sqrt{2} T^{*} \l^{(asym)} \psi_{T}
+ i \sqrt{2} {\overline{\psi}}_{T}{ \overline{\lambda}}^{(asym)}  T
\right.
\nonumber \\
& &
- ( \cD_{m} {\tilde{T}} ) (\cD^{m} {\tilde{T}} )^{*}- i
 {\overline{\psi}}_{{\tilde{T}}}   \overline{\s}^{m}
\cD_{m} \psi_{{\tilde{T}}}
- i \sqrt{2} {\tilde{T}}^{*}  \l^{(asym)} \psi_{{\tilde{T}}}
+ i \sqrt{2} {\overline{\psi}}_{{\tilde{T}}}
{ \overline{\lambda}}^{(asym)}  {\tilde{T}}
\nonumber \\
& &
-2 (\Phi^{*}_{(asym)} T^{*}) (\Phi_{(asym)} T)
-2 ( \tilde{T} \Phi_{(asym)}) (\tilde{T}^{*} \Phi^{*}_{(asym)} )\nonumber \\
& &
- \sqrt{2} ( \psi_{{\tilde{T}}} {\psi}^{(asym)} T +  \tilde{T} {\psi}^{(asym)}
 \psi_T         + \psi_{{\tilde{T}}} \Phi_{(asym)} \psi_T)
\nonumber\\
& &
- \sqrt{2} ( \overline{\psi}_{T} {\overline{\psi}}^{(asym)} \tilde{T}^{*}
+  T^{*} {\overline{\psi}}^{(asym)} \overline{\psi}_{\tilde{T}} + \psi_{T}
 \Phi_{(asym)}^{*} \overline{\psi}_{\tilde{T}} )
\nonumber \\
& &
\left.
+ \sqrt{2} \tilde{T} F_{\Phi}^{(asym)} T
+ \sqrt{2} \tilde{T}^{*} F_{\Phi}^{* (asym)} T^{*}
+ \tilde{T} D^{(asym)} T
+ \tilde{T}^{*} D^{(asym)} T^{*}
\right\}
\eeqn


\beqn
S_{fund}
& = &
+ \frac{1}{g^2} \sum_{f=1}^{n_f}
 (
- ( \cD_{m} Q_{(f)} )^{*} (\cD^{m} Q_{(f)} )
- i {\overline{\psi}}_{Q (f)}  \overline{\s}^{m}
\cD_{m} \psi_{Q (f)}
+ i \sqrt{2} Q_{(f)}^{*} \l \psi_{Q (f)}
- i \sqrt{2} {\overline{\psi}}_{Q (f)}  \overline{\lambda} Q_{(f)}
)
\nonumber \\
& &
+ \frac{1}{g^2} \sum_{f=1}^ {n_f}(
- ( \cD_{m} {\tilde{Q}}_{(f)} ) (\cD^{m} {\tilde{Q}}_{(f)} )^{*}-
 i {\overline{\psi}}_{{\tilde{Q}} (f)}  \overline{\s}^{m}
\cD_{m} \psi_{{\tilde{Q}} (f)}
- i \sqrt{2} \tilde{Q}_{(f)}  \l \psi_{{\tilde{Q}} (f)}
+ i \sqrt{2} {\overline{\psi}}_{{\tilde{Q}} (f)}
  \overline{\lambda} {\tilde{Q}}_{(f)}^{*}
)
\nonumber \\
& & + \frac{1}{g^2}  \sum_{f=1}^{n_f} (
Q^{*}_{(f)} D Q_{(f)}+{\tilde{Q}}_{(f)} D {\tilde{Q}}^{*}_{(f)} )
\nonumber \\
& & +
\frac{1}{g^2}  \sum_{f=1}^{n_f} \{
 - (m_{(f)})^2 ( Q^{*}_{(f)} Q_{(f)}+{\tilde{Q}}_{(f)} {\tilde{Q}}^{*}_{(f)})
-  m_{(f)} ( {\tilde{\psi}}_{Q {(f)}} \psi_{Q {(f)}}
+ \bar{{\tilde{\psi}}}_{Q {(f)}} \psib_{Q {(f)}} )
\nonumber\\
& &
- \sqrt{2}
(
Q^{*}_{(f)} \Phi^{\dagger} Q_{(f)} + \tilde{Q}_{(f)} \Phi^{\dagger}
 \tilde{Q}^{*}+ Q_{(f)}^{*} \Phi Q_{(f)} + \tilde{Q}_{(f)} \Phi
 \tilde{Q}_{(f)}^{*}
)
\nonumber\\
& &
-2 Q_{(f)}^{*} \Phi^{\dagger} \Phi Q_{(f)} -2 \tilde{Q}_{(f)} \Phi^{\dagger}
 \Phi \tilde{Q}_{(f)}^{*}
\nonumber\\
& &
- \sqrt{2} ( \psi_{{\tilde{Q}}(f)} {\psi} Q_{(f)}+  \tilde{Q}_{(f)}  {\psi}
 \psi_{Q (f)}           + \psi_{{\tilde{Q}}(f)} \Phi \psi_{Q (f)})
\nonumber\\
& &
- \sqrt{2} ( \overline{\psi}_{Q (f)} {\overline{\psi}} \tilde{Q}_{(f)} ^{*}
+  Q_{(f)} ^{*} {\overline{\psi}} \overline{\psi}_{\tilde{Q} (f)}
        + \psi_{Q (f)} \Phi^{\dagger} \overline{\psi}_{\tilde{Q} (f)} )
\nonumber\\
& &
+ \sqrt{2} \tilde{Q}_{(f)}  F_{\Phi} Q_{(f)}+ \sqrt{2}
\tilde{Q}_{(f)} ^{*} F_{\Phi}^{\dagger} Q^{*}_{(f)}
\}
\eeqn
where
\beqn
D_i^{\;\;\; j} &=& [\Phi^{\dagger} , \Phi]_i^{\;\;\; j}
        + \sum_{f=1}^{n_f}      (
Q^{* \; j}_{(f)}  Q_{(f) \; i}  +{\tilde{Q}}^j_{(f)}
  {\tilde{Q}}^{*}_{(f) \; i}    )
        +2 T^{* \; jk} T_{ki} + 2 \tilde{T}^{jk} \tilde{T}^{*}_{ki}
\label{eqn;D}
\\
F_{\Phi \; i}^{\;\;\;\;\;\; j} &=&
        - \sum_{f=1}^{n_f}      (
        \sqrt{2} Q^{* \; j}_{(f)}  \tilde{Q}^{*}_{(f) \; i}     )
        - \sqrt{2} T^{* \; jk} T^{*}_{ki}
\label{eqn;FPhi}
\eeqn
  Here ${\cal D}_{m} = i v_{m}$ with $v_{m}$ in  appropriate representations.
 $\Phi_{(asym)}$ , $\psi^{(asym)}$ and $F_{\Phi}^{(asym)}$ are the fields in
 anti-symmetric representation.

  Let us now find  a relationship
 of $ S_{{\rm vec}} +S_{{\rm asym}}$  in eq.~(\ref{eq:def}) with the reduced
 action of the four dimensional
 ${\cal N} =4$ supersymmetric Yang-Mills written again in terms of
 superfields. This latter action in turn
  is related in the component form to the reduced action of the
    ten-dimensional ${\cal N} =1$ Yang-Mills, which is  nothing but
 the type $IIB$ matrix model \cite{KEK}.

 First note that $ S_{ {\rm vec}} +S_{ {\rm asym}}$ in eq.~
(\ref{eq:def})  is written as
\beqn
\label{eq:vvaym}
 S_{{\rm vec}} +S_{{\rm asym}} \equiv S_{ {\rm adj} + {\rm asym}} &=&
 \frac{1}{4 g^{2}} \;  Tr \left(
 \int d^2 \th W^{\a} W_{\a} + h.c. +
4 \int d^2 \th d^2 \thb \Phi^{\dg i} e^{2V} \Phi_{i} e^{-2V} \right)
 \nonumber   \\
  &+&  \frac{1}{\sqrt{2}g^{2}} \;  Tr \left(
\int d^2 \th d^2 \thb
 \epsilon^{ijk} \left[  \Phi_{i} , \left[ \Phi_{j},  \Phi_{k} \right]
\right]  + h.c.  \;\;\right) \;\;\;,
\eeqn
  where  we have introduced the notation
\beqn
  \Phi_{1} \equiv  \Phi \;,\;\;  \Phi_{2} \equiv Y\;,\;\;
 \Phi_{3}  \equiv  \tilde{Y} \;\;\;.
\eeqn
  The form of eq.~{(\ref{eq:vvaym})  is nothing but  the reduced action
 of $d=4$, ${\cal N}=4$ super Yang-Mills, which we denote by 
 $S_{{\cal N}=4}^{~d=4}$ :
\beq
S_{ {\rm adj} + {\rm asym}} = S_{{\cal N}=4}^{~d=4} \;\;.
\eeq

  It is expedient to introduce  the projector acting on $U(2k)$ matrices:
\beqn
 \hat{\rho}_{\mp} \bullet  \equiv \frac{1}{2} \left( \bullet \mp F^{-1}
 \bullet^{t}  F \right) \;\;\;.
\eeqn
 The action  of  $\hat{\rho}_{-}$   and that of  $\hat{\rho}_{+}$   take
 any $U(2k)$ matrix  into  the matrix lying in the adjoint representation
 of $USp(2k)$  and that in the antisymmetric representation respectively.
   We can  therefore write
\beqn
 V =  \hat{\rho}_{-} \ud{V} \;,\;\;   \Phi_{1} = \hat{\rho}_{-}
 \ud{\Phi_{1}} \;,\;\;
   \Phi_{i}  =   \hat{\rho}_{+} \ud{\Phi_{i}} \;, \;\;  i = 2,3 \;\;,
\eeqn
 where the  symbols with underlines  lie in the adjoint representation
 of $U(2k)$.

We now invoke the well-known fact that the action of $d=4$, ${\cal N}=4$
 super Yang-Mills  can be obtained from the dimensional reduction of
 $d=10$, ${\cal N}=1$ super Yang-Mills down to four dimensions \cite{BSS}.
 This is stated as
\beqn
\label{eq:n4d4d10}
 S_{{\cal N}=4}^{~d=4} ( \ud{v}_{m} , \ud{\Phi}_{i},
 \ud{\lambda}, \ud{\psi}_{i},
\bar{\ud{\Phi}}_{i}, \bar{\ud{\lambda}},  \bar{\ud{\psi}}_{i}, )  &=&
   S^{d=10}_{{\cal N}=1}(\ud{v}_{M},  \ud{\Psi} ),   \nonumber \\
   S^{d=10}_{{\cal N}=1}(\ud{v}_{M},  \ud{\Psi} )  &=&
  \frac{1}{g^{2}} Tr \left( \frac{1}{4} \left[\ud{v}_{M},
 \ud{v}_{N} \right]
\left[ \ud{v}^{M}, \ud{v}^{N} \right] - \frac{1}{2}
\bar{\ud{\Psi}} \Gamma^{M} \left[ \ud{v}_{M},
\ud{\Psi} \right] \right) \;.
\eeqn
 Here
\beqn
\label{eq:phiv}
\Phi_{i} &=& \frac{1}{\sqrt{2}} \left( v_{3+i} +i v_{{6+i}}
 \right) \;\;, \;  \nonumber \\   {\rm and} \;\;\;\;\;\;\;
\Psi  &=&  \left(
 \lambda , 0, \psi_{1}, 0 ,\psi_{2}, 0  ,  \psi_{3}, 0 ,
  0, \bar{\lambda}, 0,
 \bar{\psi}_{1}, 0,
 \bar{\psi}_{2},  0, \bar{\psi}_{3} \right)^t \;\;,
\eeqn
   which is a thirty-two component Majorana-Weyl spinor
satisfying
\beqn
  C \bar{\Psi}^t = \Psi  \;
 , \; \Gamma_{11} \Psi =  \Psi \;\;\;.
\label{eq:majorana-weyl con}
\eeqn
With regard to eqs. (\ref{eq:phiv}) and (\ref{eq:majorana-weyl con}),
the same is true for objects with underlines.
 The ten dimensional gamma matrices have been denoted by 
 $\Gamma^{M}$. We will not spell out their explicit form which is determined
 from eqs.~(\ref{eq:n4d4d10}),(\ref{eq:phiv}).

What we have established through the argument above are
summarized as the following formulas useful in later sections. 
\beqn
 S_{{\rm adj}+{\rm asym}} =  S_{{\cal N}=1}^{d=10}(
 \hat{\rho}_{b\mp}\ud{v}_{M}, \hat{\rho}_{f\mp}\ud{\Psi}  ) \;\;,
\label{eq:ours=10dSYM}
\eeqn
  where $\hat{\rho}_{b\mp} $ is a matrix with Lorentz indices
and  $\hat{\rho}_{f\mp} $ is a matrix with spinor indices.
\beqn
 \hat{\rho}_{b\mp}
 &=& diag
(\hat{\rho}_{-},\hat{\rho}_{-},\hat{\rho}_{-},
 \hat{\rho}_{-},\hat{\rho}_{-},\hat{\rho}_{+},
 \hat{\rho}_{+},\hat{\rho}_{-}, \hat{\rho}_{+},\hat{\rho}_{+} )
\nonumber \\
 \hat{\rho}_{f\mp}   &=&
\hat{\rho}_{-} 1_{(4)} \otimes
\left( \begin{array}{cccc}
        1_{(2)}& & & \\
               & 0 & & \\
               &   & 1_{(2)} & \\
               &   &         &0
        \end{array}
\right)
+
\hat{\rho}_{+} 1_{(4)} \otimes
\left( \begin{array}{cccc}
        0& & & \\
               & 1_{(2)} & & \\
               &   & 0 & \\
               &   &         & 1_{(2)}
        \end{array}
\right)
 \;\;.
\label{eq:projectors of b f}
\eeqn

 The notable properties of the model  discussed in \cite{IT} are,
among other things,\\
1) it possesses eight dynamical and eight kinematical supersymmetries.\\
2) translations in six out of ten directions are broken.\\
We will discuss implication of these in subsequent sections.

\section{$USp(2k)$ planar diagrams, matrix twist and matrix T dual}

  We now discuss  $USp(2k)$ planar diagrams to see how they create
 nonorientable surfaces approximated by the dense set of Feynman diagrams.
 We set aside the fields lying in the fundamental representation in this
 section.
We ignore fermions as well.
It is well-known that the large $k$ expansion of ordinary  $U(2k)$ pure
 Yang-Mills theory in arbitrary dimensions is a topological (genus) expansion
 of the two-dimensional (discretized) surfaces created by the Feynman
 diagrams \cite{tH}.  
 It is simple to see  how  this is modified by
 $USp(2k)$ Feynman diagrams where some of them are in the adjoint while
 the others are in the nonadjoint (antisymmetric).

  Recall that the propagator in the $U(2k)$ gauge theory is
\beqn
\label{eq:u2k}
 \langle \ud{v}_{m\;r}^{\;\;\;\;\;s} \ud{v}_{n\;i}^{\;\;\;\;\;j} \rangle  =
  \delta_{i}^{\;s} \delta_{r}^{\;j} \delta_{mn} D \;\;  = \mbox{figure 1}
\eeqn
\bfig[ht]
\epsfysize=2cm
 \begin{center}
\epsfbox{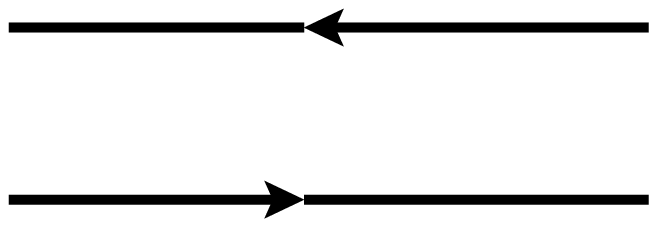}
\caption{propagator}
 \label{fig1}
\end{center}
\efig
  From now on we ignore the $D$ function as its dependence on the
 arguments  is irrelevant to the present discussion.
 The three and four point vertices are depicted in figure 2 .
\bfig[ht]
\epsfysize=4cm
 \begin{center}
\epsfbox{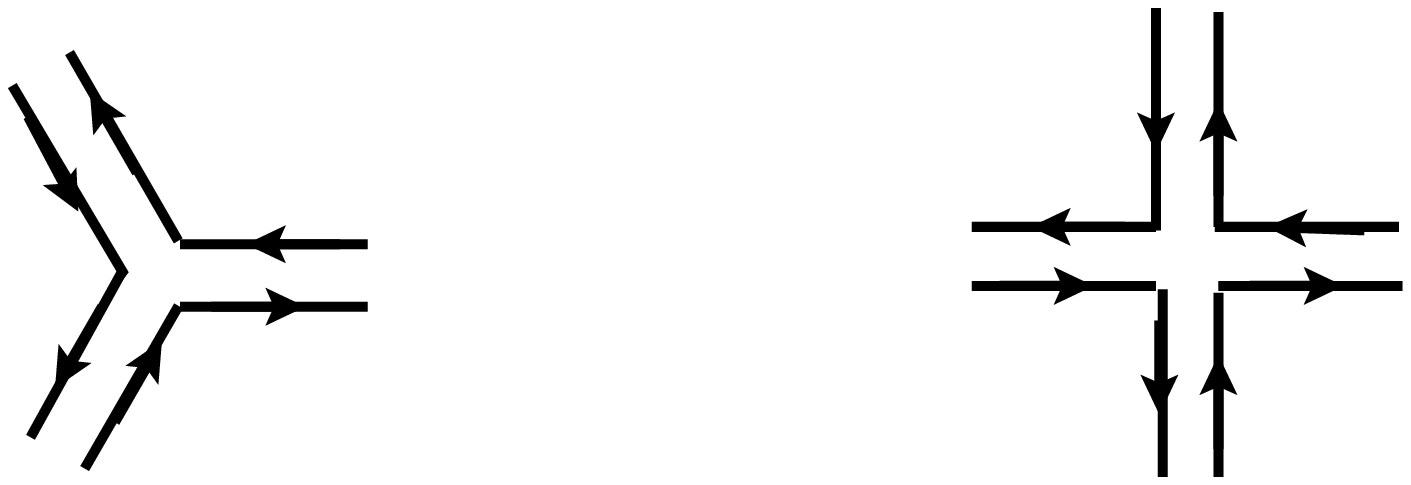}
\caption{three point vertex and four point vertex}
 \label{fig2}
\end{center}
\efig

Let $\ud{{\cal G}}$ be a $U(2k)$ Feynman diagram. Its dependence on
$g^{2}$ and on $k$  denoted by $r\left( \ud{{\cal G} } \right)$ is known to be
\beqn
  r \left( \ud{{\cal G} } \right)
&=& \left( g^{2}k \right)^{{\cal E}-{\cal V}} k^{\chi} \;\;,
\nonumber \\
 \chi &=& {\cal F} -{\cal E} + {\cal V} =  2- 2{\cal H}\;\;,
\eeqn
  where ${\cal E}$ is the number of external lines in $\ud{ {\cal G} }$,
 which
 is  also the number of edges of the surface, 
 while ${\cal V}$ is the number of
 three and four point vertices in $\ud{ {\cal G} }$ and  is on the surface.
 The number of faces or index loops and the number of holes of the surface
 are denoted by ${\cal F}$ and by ${\cal H}$ respectively.

  In $USp$ Feynman diagrams, eq.~(\ref{eq:u2k}) is modified to
\beqn
\label{eq:prop}
 \langle v_{M\;r}^{\;\;\;\;\;s} v_{N\;i}^{\;\;\;\;\;j} \rangle  &=&
 \sum_{a=1}^{2k^{2}\pm k}   \left(t^{a}\right)_{r}^{\;s}
 \left(t^{a}\right)_{i}^{\;j}
\delta_{MN}  \nonumber \\
&=& \left( \hat{\rho}_{\mp} \right)_{r\;i}^{\;s\;j} \delta_{MN} \nonumber \\
&=& \mbox{figure 3}
\eeqn
  Here we have treated  the  adjoint and nonadjoint cases collectively.
  Similarly, let  ${\cal G}$ be a  $USp$ Feynman  diagram. As the propagator
  contains the second term which reduces the number of index loops by one,
 $r({\cal G})$ depends upon how many times double lines representing
 propagators cross. Clearly
\beqn
r({\cal G}) = r({\cal G}; c) =   \left( g^{2}k \right)^{{\cal E}-{\cal V}} k^{\chi -c}\;\;,
\eeqn
where $c$ denotes the number of crossings.
\bfig[ht]
\epsfysize=2cm
 \begin{center}
\epsfbox{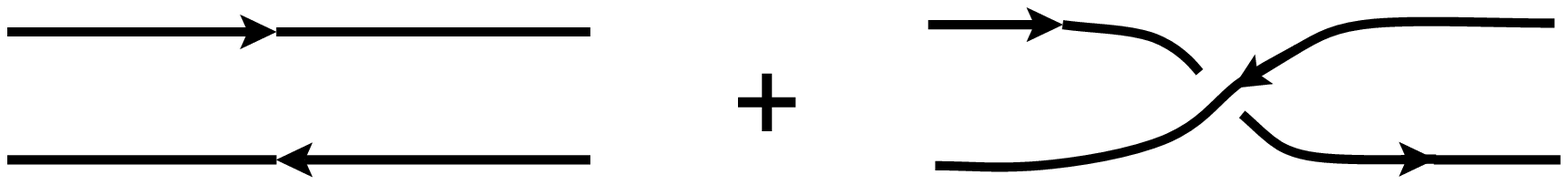}
\caption{propagator in $USp(2k)$}
 \label{fig3}
\end{center}
\efig

   We still need to show that $c$ denotes the number of cross caps 
  and not the number of boundaries. Let us recall that, according to
 the present point of view, the two-dimensional surface swept by a string is
  formed by the dense set of Feynman diagrams.
  To render this more tangible and more than a combinatorial argument, we note
  that, via the Schild gauge correspondence, the algebra acting on the
 functions on the string world sheet must be isomorphic to the large
 $k$ limit of the appropriate Lie algebra acting on matrices.
   For this, it is enough to adopt the argument of Pope and
 Romans\cite{PR} on area-preserving
 diffeomorphisms on $RP^{2}$ and 
 the large $k$ limit of the $usp(2k)$ Lie algebra in the
 present context.
  Consider first the sphere parametrized by three coordinates $x^{i},$
 $i=1,2,3$  such that $x^{i}x^{i} =1$.  The complete set of functions on the
 sphere is the spherical harmonics  represented by
\beqn
   Y^{(p)}(x^{i}) =  a_{i_{1}, \cdots i_{p}} x^{ i_{1}} \cdots
 x^{i_{p}} \;\;,
\eeqn
  where $a_{i_{1}, \cdots i_{p}}$  are totally symmetric and traceless
  constants.
The algebra of area preserving diffeomorphisms is  defined by a bracket of
 two functions $A(x^{i})$ and $B(x^{i})$:
\beqn
\label{eq:APD}
 \left\{ A, B \right\} \equiv \epsilon_{ijk} x^{i} \partial_{j}A
 \partial_{k}B \;\;.
\eeqn
  When $A= Y^{(m)}$, $B= Y^{(n)}$, a finite sum of irreducible polynomials
 $Y^{(p)}$, $ \mid m-n\mid \leq p \leq m+n-1$ is generated.
  This algebraic structure is obtained by the large $k$ limit of
  the $su(2k)$ Lie algebra in the form of maximal $su(2)$ embeddings:
\beqn
  \Lambda^{(p)} =  a_{i_{1}\cdots i_{p}} \Sigma^{ i_{1}} \cdots
 \Sigma^{i_{p}} \;\;,   p= 1 \sim k-1.
\eeqn
  Here, $\Sigma^{i}$ are the $su(2)$ generators in the 
 $2k$-dimensional representation.
  On the other hand,  $RP^{2}$ geometry is obtained from the sphere by
  the antipodal identification $x^{i} \rightarrow -x^{i}$, under which
  the harmonics splits into even and odd ones.
 Only the odd ones are responsible for forming the algebra of
 area-preserving diffeomorphism on $RP^{2}$: this is clear from
 eq.~(\ref{eq:APD}).
 We see that the diffeomorphisms of $RP^{2}$ are generated by the large $k$
 limit of the generators
\beqn
\label{eq:oddL}
  \Lambda^{(2p-1)} =  a_{i_{1}\cdots i_{2p-1}} \Sigma^{ i_{1}} \cdots
 \Sigma^{i_{2p-1}} \;\;,   p= 1, 2, \cdots k\;\;.
\eeqn
  As shown by Pope and Romans, the algebra formed by eq.~(\ref{eq:oddL}) is
 the Lie algebra $usp$. 
This concludes that the diagrams generated from the propagator 
( eq.~(\ref{eq:prop})) and the vertices contain $RP^{2}$. The theory we are
 constructing via matrices is the reduce model of nonorientable  strings.

  To extend the above argument to higher genera with crosscaps, let us note
that the role of the matrix $F$  can be seen by the correspondence with
the twist operation in the operator formalism  of the first quantized string.
Ten of the noncommuting coordinates $v_{M}$, which are dynamical variables,
satisfy
\beqn
\label{eq:rel}
v_{i}^{t} &=& - F v_{i} F^{-1}\;\; , \;\; i \in \{\{ 0, 1, 2, 3, 4, 7 \}\}
  \equiv {\cal M}_{-} \;\;, \nonumber \\
v_{I}^{t} &=& \; F v_{I} F^{-1} \;\; , \;\; I \in
 \{\{ 5, 6, 8, 9 \}\} \equiv {\cal M}_{+} \;\;.
\eeqn
 The $v_{M}$ are noncommuting counterparts of the ten string coordinates
 $X_{M}$. That this is more than just an analogy is clear as the limit exists
 from our action to the string action of the Schild type gauge.
  Taking the transpose is interpreted as flipping the direction of
 an arrow drawn on a string.  The operation $F$ is
 the matrix analog of the twist
 operation $\Omega$ \footnote{In the context of ref. \cite{BFSS},
 see ref. \cite{Mtwist}.}.
The classical counterpart of eq.~(\ref{eq:rel}) is therefore
\beqn
\label{eq:clrel}
X_{i}(\bar{z}, z) &=& - \Omega X_{i} (z, \bar{z}) \Omega^{-1} \;\; , \;\;
 i \in {\cal M}_{-} 
   \nonumber \\
X_{I}(\bar{z}, z) &=& \;  \Omega X_{I}( z,\bar{z}) \Omega^{-1} \;\; , \;\;
 I \in {\cal M}_{+}  \;\;.
\eeqn
 The presence of four dimensional fixed surfaces
 (orientifold surfaces, O3s)
 becomes clear from this equation (\ref{eq:clrel}).
 We conclude that our model is a matrix model on a large volume $T^{6}/Z^{2}$
 orientifold. This is consistent with that the translations
 in six out of ten directions are broken.

 The $T$ duality transformation plays an interesting role in matrix models
 as it relates worldvolume theories of various dimensions via Fourier
 transforms. We will now find how the matrix $T$ dual behaves under $F$.
 First, let us impose periodicities  with period $2\pi R$  for $L$
 out of the ten coordinates.  Recall that
\beqn
 Y_{\ell} \equiv  \hat{T}\left[ X_{\ell}\right] 
 &\equiv&  X_{\ell\;R}-  X_{\ell\;L} \;\;,   
\\
\hat{T}\left[ X_{\ell}\right](\bar{z}, z) 
&=&
\left\{
\begin{array}{c}
 + \Omega \; \hat{T}\left[ X_{\ell}\right] (z, \bar{z}) \; \Omega^{-1} 
 \;\; {\rm if }\; \ell \in {\cal M}_{-} 
\\
 - \Omega \;  \hat{T}\left[ X_{\ell}\right] (z, \bar{z}) \; \Omega^{-1}  
\;\; {\rm if }\; \ell \in {\cal M}_{+} 
\end{array}
\right. \;\; .
\label{eq:string T-dual flip}
\eeqn

  To impose periodicities on infinite size matrices $v_{\ell}$,
  we decompose  $v_{\ell}$ into  blocks of $n \times n$  matrices.
  Specify each individual block by an $L$-dimensional row vector
 $\vec{a}$ and an $L$-dimensional column vector $\vec{b}$:
 $\left( v_{\ell}\right)_{\vec{a}, \vec{b}} \equiv  \sqrt{\alpha^{\prime}}
 \langle \vec{a} \mid  \hat{v}_{\ell} \mid \vec{b} \rangle $.
  Let the shift vector be
\beqn
 \left( U(i)\right)_{\vec{a}, \vec{b}} =  \left( \prod_{j (\neq i)}
 \delta_{a_{j}, b_{j}} \right) \delta_{a_{i}, b_{i}+1} \;\;.
\eeqn 

 The condition to be imposed is
\beqn
\label{eq:period}
U(i) v_{\ell} U(i)^{-1} &=&  v_{\ell} - \delta_{\ell,i}
 R /\sqrt{\alpha^{\prime}} \;\;.
\eeqn
 The solution in the Fourier transformed space is
\beqn
\langle  \vec{x} \mid \hat{v}_{\ell} \mid \vec{x}^{\prime} \rangle
 &=& -i \left( \frac{\partial}{\partial x^{\ell}} +i \tilde{v}_{\ell}(\vec{x})
 \right)
  \delta^{(L)} \left( \vec{x}- \vec{x}^{\prime} \right)
 \;\;, \\
\tilde{v}_{\ell}(\vec{x}) &=&  \sum_{\vec{k} \in {\cal Z}^{L}}
 \tilde{\tilde{v}}_{\ell} (\vec{k})
  \exp \left( \frac{-i \vec{k} \cdot \vec{x}}{\tilde{R}} \right)
 \;\;, \nonumber \\
 \tilde{R} &\equiv& \alpha^{\prime} /R \;\;\;.
\eeqn
The matrix $T$ dual is nothing but the Fourier transform: it interchanges the
  radius parameter $R$  setting the period of the original matrix index
 with the dual radius $\tilde{R}$  which is the period of the space
  Fourier conjugate to the matrix index.
 Let us write
\beqn
  \hat{T} \left[ \left( v_{\ell}\right)_{\vec{a}, \vec{b}} \right]
\equiv \langle \vec{x} \mid \hat{v}_{\ell} \mid \vec{x}^{\prime} \rangle \;\;.
\eeqn 
  Multiply eq.~(\ref{eq:rel}) written in the bracket notation
\beqn
 \langle \vec{d} \mid \hat{v}_{\ell} \mid \vec{a} \rangle
=
\mp  \sum_{\vec{b}, \vec{c}} \langle \vec{a}\mid \hat{F} \mid \vec{b} \rangle
\langle \vec{b} \mid \hat{v}_{\ell} \mid \vec{c} \rangle \langle
 \vec{c} \mid \hat{F}^{-1} \mid \vec{d} \rangle
 \;\;
\eeqn
by $\langle \vec{x} \mid \vec{a} \rangle \langle \vec{d} \mid
 \vec{x^{\prime}} \rangle$  $= \langle \vec{a} \mid \vec{x} \rangle^{*} 
\langle \vec{x^{\prime}} \mid \vec{d} \rangle^{*}$.
  Sum over $\vec{a}$ and  $\vec{d}$.
  From  the left hand side, we obtain
\beqn
 - \left[ -i \frac{\partial}{\partial x^{\prime \ell}} -
 \tilde{v}_{\ell} ( - \vec{x}^{\prime}) \right] 
\delta^{(L)} (  \vec{x}^{\prime} - \vec{x}) \;\;.
\eeqn
  We find
\beqn
 \hat{T} \left[ v_{\ell} \right]^{t} 
=
\left\{
\begin{array}{c}
+   \hat{T} \left[ F \right]  \hat{T} \left[ v_{\ell} \right]
 \hat{T} \left[ F^{-1} \right]
 \;\; {\rm if }\; \ell \in {\cal M}_{-} \\
-  \hat{T} \left[ F \right]  \hat{T} \left[ v_{\ell} \right]
 \hat{T} \left[ F^{-1} \right]
 \;\; {\rm if }\; \ell \in {\cal M}_{+} 
\end{array}
\;\;, \right.
\label{eq:matrix T-dual flip}
\eeqn
  provided 
\beqn
 \tilde{v}_{\ell} ( - \vec{x}^{\prime}) =
- \tilde{v}_{\ell} (  \vec{x}^{\prime})
\;\;.
\label{eq:odd function}
\eeqn
  It is satisfying to see
 that the sign change of eq. (\ref{eq:matrix T-dual flip})
 from eq. (\ref{eq:rel}) under
 the matrix $T$ dual is in accordance with the sign change of 
 eq. (\ref{eq:string T-dual flip}) from eq. (\ref{eq:clrel})

  One can now imagine imposing periodicities 
 with periods depending on the directions
 and  letting some of the radii zero.  The $T$ duality provides
 worldvolume gauge theories in various dimensions.  We will discuss a few
 cases later.

\section{$USp$ projector and supersymmetry}

  We will now derive a set of conditions under which  the projectors
  $\hat{\rho}_{b \mp}$, $\hat{\rho}_{f \mp}$,
 which act respectively
 on $\ud{v}_{M}$ and $\ud{\Psi}$,
 and dynamical $\delta^{(1)}$ as well as
 kinematical $\delta^{(2)}$ supersymmetry commute.
Our choice for
  $\hat{\rho}_{b \mp}$ and  that for $\hat{\rho}_{f \mp}$ emerge as  the case
which passes the tight constraint of having eight dynamical and eight
 kinematical supersymmetries. 
 Let us start with
\beqn
   \delta^{(1)} \ud{v}_{M} &=& i \bar{\epsilon} \Gamma_{M} \ud{\Psi}
\label{eq:d11} \\
\delta^{(1)} \ud{\Psi} &=& \frac{i}{2} \left[ \ud{v}_{M}, \ud{v}_{N} \right]
 \Gamma^{MN} \epsilon  \label{eq:d12} \\
 \delta^{(2)}\ud{v}_{M} &=& 0 \label{eq:d21} \\
\delta^{(2)} \ud{\Psi} &=& \xi \;\;\;.  \label{eq:d22}
\eeqn
 Let us write generically
\beqn
\label{eq:prv}
 v_{M} &\equiv&  \delta_{M}^{~N} \hat{\rho}_{b \mp}^{(N)}
 \ud{v}_{N} \;\;\;  \nonumber \\
 \Psi_{A} &\equiv&  \delta_{AB} \hat{\rho}_{f \mp}^{(B)} \ud{\Psi}_B \;\;.
\eeqn
  The condition
 $\left[\hat{\rho}_{b \mp}, \delta^{(1)} \right] \ud{v}_{M} =0$  together
 with eq.~(\ref{eq:d11}) gives
\beqn
\label{eq:1}
 \sum_{A=1}^{32} \left( \bar{\epsilon} \Gamma_{M}\right)_{A} \left(
\hat{\rho}_{f \mp}^{(A)}-  \hat{\rho}_{b \mp}^{(M)} \right)  \ud{\Psi}_{A}
= 0 \;\;,
\eeqn
  with index $M$ not summed.
  The condition
\beqn
\label{eq:con2}
\left.
\left[\hat{\rho}_{f \mp}, \delta^{(1)} \right] \ud{\Psi}
\right|_{\ud{v}_{M}
\rightarrow \hat{\rho}_{b \mp}\ud{v}_{M} } =0
\eeqn
 together with eq.~(\ref{eq:d12}) provides
\beqn
\label{eq:2}
 \left( 1 - \hat{\rho}_{f \mp}^{(A)} \right) \left[ \hat{\rho}_{b \mp}^{(M)}
 \ud{v}_{M},  \hat{\rho}_{b \mp}^{(N)} \ud{v}_{N} \right]
  \left( \Gamma^{MN} \epsilon \right)_{A} =0  \;\;\;.
\eeqn
  The restriction  at eq.~(\ref{eq:con2}) comes from the fact that
 eq.~(\ref{eq:d12}) is true only on shell.
Eq. ~(\ref{eq:d21})  does not give us anything new  while
$\left[\hat{\rho}_{f \mp}, \delta^{(2)} \right] \ud{\Psi} =0$ with
eq.~(\ref{eq:d22}) gives
\beqn
 \xi_{A} 1 = \xi_{A} \hat{\rho}_{f \mp}^{(A)} 1 \;\;\;,
\label{eq: condition kin susy}
\eeqn
  with index $A$ not summed.

  In order to proceed further, we rewrite  eq.~(\ref{eq:prv}) explicitly as
\beqn
 \hat{\rho}_{b \mp}^{(M)}  &\equiv& \Theta (M\in {\cal M}_{-}) \hat{\rho}_{-}
 +  \Theta (M\in {\cal M}_{+}) \hat{\rho}_{+}  \nonumber \\
\hat{\rho}_{f \mp}^{(A)} &\equiv&  \Theta (A\in {\cal A}_{-}) \hat{\rho}_{-}
+  \Theta (A\in {\cal A}_{+}) \hat{\rho}_{+} \;\;\; ,
\label{eq: explicit proj}
\eeqn
where
\beq
 {\cal M}_{-} \cup {\cal M}_{+}
 =  \{ \{ \; 0,1,2,3,4,5,6,7,8, 9 \; \} \} \;\; ,
\;\;
  {\cal M}_{-} \cap {\cal M}_{+} = \phi \;\;,
\label{eq: M condition}
\eeq
\beq
 {\cal A}_{-} \cup {\cal A}_{+}
 = \{ \{ \; 1,2,5,6,9,10,13,14,19,20,23,24,27,28,31,32 \; \} \} \;\; ,
\;\;
  {\cal A}_{-}\cap  {\cal A}_{+}=\phi \; .
\label{eq: A condition}
\eeq
  We find that  eq.~(\ref{eq:1}) gives
\beqn
\label{eq:set1}
  \left( \bar{\epsilon} \Gamma_{M_{-}} \right)_{A_{+}} = \left( \bar{\epsilon}
 \Gamma_{M_{+}} \right)_{A_{-}} =0 \;\;,
\eeqn
     while eq.~(\ref{eq:2}) gives
\beqn
\label{eq:set2}
  \left( \Gamma^{M_{-}N_{+}} \epsilon \right)_{A_{-}} &=& 0 \nonumber \\
 \left( \Gamma^{M_{-}N_{-}} \epsilon \right)_{A_{+}}  &=&
\left( \Gamma^{M_{+}N_{+}} \epsilon  \right)_{A_{+}} =0 \;\;.
\eeqn
Equation (\ref{eq: condition kin susy}) gives
\beqn
\xi_{A_{-}} = 0 \;\;.
\eeqn
 As  we consider the case of eight kinematical
supersymmetries, the number of elements of the sets
  denoted by $\mbox{\boldmath $\sharp$} ({\cal A}_{\pm})$  must be
\beq
\mbox{\boldmath $\sharp$} ({\cal A_{-}}) = 8 \;\; \mbox{ and } \;\;  
\mbox{\boldmath $\sharp$}({\cal A_{+}}) = 8
\label{eq:number of A+A-} \;.
\eeq

  Eqs.~(\ref{eq:set1}) and (\ref{eq:set2})  are regarded as the ones which
determine the anticommuting parameter $\e$, and
the sets  ${\cal A}_{+}$, ${\cal A}_{-}$, ${\cal M}_{+}$ and
${\cal M}_{-}$. In addition they must satisfy the conditions 
(\ref{eq: M condition}), (\ref{eq: A condition}) and
(\ref{eq:number of A+A-}).
 
 We search for solutions by first trying out as an input an appropriate
 thirty-two component anticommuting parameter $\e$
satisfying Majorana-Weyl condition.

Given $\e$, we  see if  we can determine ${\cal A}_{+}$, ${\cal A}_{-}$,
 ${\cal M}_{+}$ and ${\cal M}_{-}$ successfully.
Our strategy is:\\

{\bf (i)}
calculate $\left( \eb \Gamma^{M} \right)_A$
and $ \left( \Gamma^{MN} \epsilon \right)_{A} $
for all $M$, $N$ and $A$.
\\

{\bf (ii)}
calculate
${\displaystyle  \sum_A  \left( \bar{\epsilon}
\Gamma_{M_{1}} \right)_{A}
\left( \bar{\epsilon} \Gamma_{M_{2}} \right)_{A}  }$ .
If this value is nonzero,
the both indices $M_1$ and $M_2$ belong to
either ${\cal M}_{-} $ or $ {\cal M}_{+}$.
We can, therefore, divide ${\cal M}_{-} \cup {\cal M}_{+}$ into two sets.\\

{\bf (iii)} from eq. (\ref{eq:set2}) we see that
if $\left( \Gamma^{M_{-} N{+}} \e \right)_A \neq 0$, 
then $A \in {\cal A}_{+}$.
  If  $\left( \Gamma^{M_{-} N{-}} \e \right)_A \neq 0$
or 
 $\left( \Gamma^{M_{+} N{+}} \e \right)_A \neq 0$, 
then $A \in {\cal A}_{-}$ .
Use the results of {\bf (i)} and {\bf (ii)} to determine ${\cal A}_{-}$ and
${\cal A}_{+}$.
We must  then check  if $\mbox{\boldmath $\sharp$} ({\cal A_{-}}) = 8 $ ,  
$\mbox{\boldmath $\sharp$}({\cal A_{+}}) = 8$
and ${\cal A}_{-} \cap  {\cal A}_{+} = \phi $.
If  these are not satisfied,  our original input $\e$ is not a solution.\\

{\bf (iv)}
  from eq. (\ref{eq:set1}) we  see that
if $\left( \eb \Gamma_{M_{-}}  \right)_A \neq 0$ then $A \in {\cal A}_{-}$
, and if  $\left(\eb \Gamma^{M_{+}}  \right)_A \neq 0$  then
$A \in {\cal A}_{+}$ .
Determine ${\cal A}_{-} $
 and $ {\cal A}_{+}$.
If ${\cal A}_{-} $ and $ {\cal A}_{+}$ determined this way are consistent with
the result from {\bf (iii)}, we obtain a solution to
eqs.~(\ref{eq:set1}) and (\ref{eq:set2}).
This also determines ${\cal M}_{-} $ and  $ {\cal M}_{+}$
 as we have two ways of choosing them from {\bf (ii)}.

  We have tried out many cases, some of which we will describe.
The case leading to our model is
\beq
\e = (\e_0, 0, \e_1, 0, 0,0,0,0, 0, \eb_0, 0 , \eb_1, 0,0,0,0  )^t \;\; .
\eeq
Note that $\e_0$, $\e_1$, $\eb_0$ and $\eb_1$ are two-component
anticommuting parameters.  From
 step {\bf (ii)}, we see
 ${\cal M}_{-} \cup {\cal M}_{+}$ are divided into two sets:
\beq
\{\{ \; 0,1,2,3,4,7 \;\}\} \; \mbox{ and } \; \{\{ \; 5,6,8,9 \;\}\} \;\;.
\label{eq:two sets}
\eeq
  From step {\bf (iii)},  we find
\beqn
{\cal A}_{-} &=&  \{ \{ \; 1,2,5,6,19,20,23,24 \; \} \} \;\; ,
\nonumber\\
{\cal A}_{+} &=&  \{ \{ \; 9,10,13,14,27,28,31,32 \; \} \} \;\; .
\label{eq:A-A+}
\eeqn
 From step {\bf (iv)}, we obtain
\beqn
{\cal M}_{-} &=& \{\{ \; 0,1,2,3,4,7 \;\}\}  \;\;,
\nonumber\\
{\cal M}_{+} &=&   \{\{ \; 5,6,8,9 \;\}\} \;\;.
\label{eq:M-M+}
\eeqn
We conclude   that
\beqn
 \hat{\rho}_{b\mp}
 &=& diag
(\hat{\rho}_{-},\hat{\rho}_{-},\hat{\rho}_{-},
 \hat{\rho}_{-},\hat{\rho}_{-},\hat{\rho}_{+},
 \hat{\rho}_{+},\hat{\rho}_{-}, \hat{\rho}_{+},\hat{\rho}_{+} )
\nonumber \\
 \hat{\rho}_{f\mp}   &=&
\hat{\rho}_{-} 1_{(4)} \otimes
\left( \begin{array}{cccc}
	1_{(2)}& & & \\
	       & 0 & & \\
	       &   & 1_{(2)} & \\
	       &   &         &0
	\end{array}
\right)
+
\hat{\rho}_{+} 1_{(4)} \otimes
\left( \begin{array}{cccc}
	0& & & \\
	       & 1_{(2)} & & \\
	       &   & 0 & \\
	       &   &         & 1_{(2)}
	\end{array}
\right)
 \;\; ,
\eeqn
which are the projectors of our model.

Among other cases, we have tried the following one:
\beq
\e = (\e_0, 0, \e_1, 0, \e_2,0,\e_3,0,
 0, \e_0, 0 , \e_1, 0,\e_2, 0,\e_3  )^t  \;\;.
\eeq
  From step {\bf (ii)}, we obtain
\beq
\{\{ \; 0,1,2,3,4,7 \;\}\} \; \mbox{ and } \; \{\{ \; 5,6,8,9 \;\}\} \;\;.
\eeq
 We find that $ {\cal A}_{-}$ and ${\cal A}_{+} $ determined  from
 step {\bf (iii)} do not satisfy $ {\cal A}_{-} \cap  {\cal A}_{+} = \phi $.

We have examined the following cases ( and their permutations) as well with
 no success:
\beqn
\e &=& (\e_0, 0, \e_1, 0, \e_2,0, \e_3,0,
 0, -\e_0, 0 , -\e_1, 0, -\e_2, 0, -\e_3  )^t  \;\; \nonumber \\
\e &=& (\e_0, 0, \e_1, 0, 0,0, \e_3, 0,
0, \eb_0, 0 , \e_1, 0,0,0,\e_3 )^t \;\;  \nonumber \\
\e &=& (\e_0, 0, \e_1, 0, 0,0, \e_3, 0,
0, \eb_0, 0 , \e_1, 0,0,0,-\e_3 )^t \;\; , \nonumber \\
\e &=& (\e_0, 0, \e_1, 0, 0,0, \e_3, 0,
0, \eb_0, 0 , -\e_1, 0,0,0,-\e_3 )^t \;\; , \nonumber \\
\e &=& (\e_0, 0, \e_1, 0, 0,0, \e_1, 0,
0, \eb_0, 0 , \eb_1, 0,0,0,\eb_1 )^t \;\; .
\eeqn

 There is, however, another solution which we have found.
Let
\beq
\e = (\e_0, 0, \e_1, 0, 0,0,0,0, 0, 0,0,0,0, \eb_2, 0 , \eb_3   )^t \;\; .
\label{eq:80susy para}
\eeq
  The consistent sets
\beqn
{\cal A}_{-} &=&  \{ \{ \; 1,2,5,6,27,28,31,32 \; \} \} \;\; ,
\nonumber\\
{\cal A}_{+} &=&  \{ \{ \; 9,10,13,14,19,20,23,24 \; \} \} \;\; ,
\label{eq:A-A+08}
\eeqn
\beqn
{\cal M}_{-} &=& \{\{ \; 4,7 \;\}\}  \;\; ,
\nonumber\\
{\cal M}_{+} &=&   \{\{ \; 0,1,2,3,5,6,8,9 \;\}\} \;\;.
\label{eq:M-M+08}
\eeqn
are obtained  from steps {\bf (i)}, {\bf (ii)}, {\bf (iii)} and {\bf (iv)}.
The projectors (\ref{eq: explicit proj}) are
\beqn
 \hat{\rho}_{b\mp}
 &=& diag
(\hat{\rho}_{+},\hat{\rho}_{+},\hat{\rho}_{+},
 \hat{\rho}_{+},\hat{\rho}_{-},\hat{\rho}_{+},
 \hat{\rho}_{+},\hat{\rho}_{-}, \hat{\rho}_{+},\hat{\rho}_{+} )
\nonumber \\
 \hat{\rho}_{f\mp}   &=&
\hat{\rho}_{-} 1_{(4)} \otimes
\left( \begin{array}{cccc}
	1_{(2)}& & & \\
	       & 0 & & \\
	       &   & 0 & \\
	       &   &         & 1_{(2)}
	\end{array}
\right)
+
\hat{\rho}_{+} 1_{(4)} \otimes
\left( \begin{array}{cccc}
	0& & & \\
	       & 1_{(2)} & & \\
	       &   & 1_{(2)} & \\
	       &   &         & 0
	\end{array}
\right)
 \;\; .
\label{eq:2-8 case projector}
\eeqn

  This is the case considered in ref. \cite{Mtwist,Ba}
 in the context  of $M$ theory compactification to
  the lightcone
 heterotic strings (with $\e_0$, $\e_1$, $\eb_2$ and $\eb_3$ in
  eq. (\ref{eq:80susy para}) all real). 

\section{The role of the fundamental representation
                 and anomaly cancellation of worldvolume theory} 

 So far, we have ignored the fields in the fundamental representation.
  These fields do not contribute  to the diagrams in spherical topology.
 They are irrelevant to the questions concerning with the spacetime coordinates.
 They create, however, disk diagrams and higher genera with boundaries
  and are responsible for creating an open string sector.
 This is in fact required as nonorientable closed strings by themselves are not
consistent.  It is well-known that the simplest way to establish the consistency
  is through the (global) cancellation of dilaton tadpoles between disk and
 $RP^{2}$ diagrams \cite{GS}, \cite{IM}, leading to the $SO(32)$ Chan-Paton
 factor. This survives toroidal compactifications with/without discrete
 projection \cite{AIKT}.  It should be that the sum of an infinite set of
 diagrams of the matrix model contributing to the disk/$RP^{2}$ geometry
 yields the string partition function of the disk/$RP^{2}$ diagram.
 The Chan-Paton trace at the boundary  corresponds to the trace with respect
 to the flavor index. 
 The $n_{f}$ should therefore be fixed by the tadpole cancellation.
 The flavor symmetry of the model is the local gauge symmetry of strings. 

 The lack of the combinatorial argument and the absence of the vertex operator 
  construction at this moment, 
 however, prevent us from proceeding to such calculation via
 matrices.
  Instead, we will examine gauge anomalies of worldvolume theories by taking
 the $T$ dual and subsequently the zero volume limit of $T^{6}/{\cal Z}^{2}$.
  In particular,  let us do  this for all six adjoint directions.
  The resulting theory is
 the six dimensional worldvolume gauge theory obeying 
eq.~(\ref{eq:odd function}) with matter in the
 antisymmetric and fundamental representation. 
This is the type $I$ superstrings
 in ten spacetime dimensions. 
 This case is  also the first nontrivial case of getting a potentially
 anomalous theory.
   In fact, by acting 
\beqn
\Gamma_{(6)} \equiv \Gamma^{0}\Gamma^{1}\Gamma^{2}\Gamma^{3}
\Gamma^{4} \Gamma^{7} 
\eeqn 
 on $\Psi$, we see that the adjoint fermions $\lambda$ and $\psi_{1}$ have
 chirality plus  while $\psi_{2,3}$ have chirality minus.  The fermions in the
 fundamental representation have chirality minus.
 The standard technology
to compute nonabelian anomalies is provided by the family's index theorem
 and the descent equations\cite{GSano} \cite{ZWZ}.
 We find that the condition  for the anomaly cancellation:
\beqn
   & &  tr_{adj}F^{4} - tr_{asym} F^{4} - n_{f} tr F^{4}  \;\; \nonumber \\
& & = (2k+8) trF^{4} + 3 \left( trF^{2}\right)^{2}  
  - \left( \left(2k-8\right)  trF^{4} + 3 \left( trF^{2}\right)^{2}
  \right)
  - n_{f} trF^{4}   \nonumber \\
& & = \left( 16-n_{f} \right) tr F^{4} =0 \;\;\;,
\eeqn
  where we have indicated the traces in the respective representations.
 The case $n_{f}=16$  is selected by the consistency of the theory.
In the case discussed in eq. (\ref{eq:2-8 case projector}),
we conclude from similar calculation that
the anomaly cancellation of the worldvolume two-dimensional
gauge theory selects sixteen complex fermions.

\section{ One-loop effective action and D-string solutions}

\subsection{one-loop effective action}

   In this subsection, we will establish a formula for the one-loop effective
 action of the $USp$ matrix model on a generic bosonic 
background\footnote{
  The solutions we will construct in the next subsection
 and in section $7$ are relevant
 only in the large $k$ limit.
 We will, therefore, ignore the fields lying
   in the fundamental  representation.}.
 Let us first find one-loop fluctuations on  a generic
classical solution of the $USp(2k)$  matrix model.
  We write
\beqn
  v_{m} &=& p_{m} + g a_{m} \;, \;\; (m=0 \sim 3) \;\;, \;\;\;
  \lambda = \chi_{0} + g \phi_{0}
 \;\; \; \nonumber \\
  v_{I} &=& p_{I} + ga_{I} \;, \;\; (I=4\sim 9) \;\;, \;\;\;\;\;
 \psi_{i} = \chi_{i} + g\phi_{i}   \;, \;\; (i = 1\sim3 )
\label{eq:background+fluc}
\eeqn
  with $(p_{m}, p_{i},  \chi_{0}, \chi_{i})$ a configuration satisfying
 equations of motion.
In order to fix the gauge invariance
  we add  the ghost and the gauge fixing term
\beqn
  S_{gfgh} = \frac{1}{2} Tr
\left(
[p_M , a^M]^2 - [p^K , b] [p_K , c]
\right) \;\;\; ,
\eeqn
where $ c $ and $ b $ are, respectively, the ghosts and the antighosts
 lying in the adjoint representation of $USp(2k)$.
  Denote by $S^{(2)}$ the part in $S_{{\rm adj}+{\rm asym}}$ which is
 quadratic in $a$ and $\phi$.
 The one-loop effective action $W_{\one-loop}$ is
\beqn
 W_{\one-loop} = -i log \int [da_{m}][da_{I}][d\phi_0]
[d\bar{\phi}_0][d\phi_{i}][d\bar{\phi}_{i}][d c] [d b]
  \exp \left[ iS^{(2)} + iS_{gfgh} \right] \;\;\;.
\eeqn
  Instead of resorting to the direct gaussian integrations of the expression
 above,  let us use  eqs. (\ref{eq:ours=10dSYM}) and
 (\ref{eq:projectors of b f}).  

In the same way as eq.(\ref{eq:background+fluc}),
we decompose $\ud{v}_{M}$ and $\ud{\Psi}$ into the backgrounds and
the quantum fluctuations.
Let us denote the fluctuations by $\ud{v}_{M}^{(fl)}$
and $\ud{\Psi}^{(fl)}$ .
Then from eq.(\ref{eq:ours=10dSYM}) we have
\beq
S^{(2)}
=
 S_{{\cal N}=1}^{d=10 \; (2)}(
 \hat{\rho}_{b\mp} \ud{v}_{M}^{(fl)}, \hat{\rho}_{f\mp}
\ud{\Psi}^{(fl)}  ) \;\;,
\eeq
where $ S_{{\cal N}=1}^{d=10 \; (2)}(
 \hat{\rho}_{b\mp} \ud{v}_{M}^{(fl)}, \hat{\rho}_{f\mp}
\ud{\Psi}^{(fl)}  )$ is the part in the action of $d=10$ ,
 ${\cal N}=1$ super
Yang-Mills
which is quadratic in the fluctuations.
  As the variables are explicitly projected  either onto $USp(2k)$ adjoint
  or onto antisymmetric   matrices, we can safely
 replace  the integration measure by  that of
 the $u(2k)$ Lie algebra valued matrices. We obtain
\beqn
\label{eq:oneloop}
 W_{\one-loop} &=& -i log \int [d\ud{v}_{M}^{(fl)}]
[d\ud{\Psi}^{(fl)}][d \ud{c}] [d \ud{b}]
  \exp \left[ iS_{{\cal N}=1}^{d=10 \; (2)}
(\hat{\rho}_{b\mp} \ud{v}_{M}^{(fl)}, \hat{\rho}_{f\mp} \ud{\Psi}^{(fl)})
+i S_{gfgh}(\hat{\rho}_{-} \ud{b} , \hat{\rho}_{-} \ud{c} ) \right] 
\;\;\; \nonumber \\
  &=& \frac{1}{2} \log \det \left( {\cal O}_{b} \hat{\rho}_{b\mp} \right)
  - \frac{1}{2} \log \det \left( {\cal O}_{f} \hat{\rho}_{f\mp} \left(
 \frac{1+ \Gamma_{11}}{2} \right) \right)
 - \log \det \left( \hat{P_K} \hat{P^K} \hat{\rho}_{-}  \right)
\eeqn
  where
\beqn
\label{eq:obof}
   {\cal O}_{b L}^{M} = - \delta_{L}^{~M} \hat{P}_{K}\hat{P}_{K} +
 2i \hat{F}_{L}^{~M}  \;\;, \;\;\;
  {\cal O}_{f} =   - \Gamma_{M} \hat{P}^{M} \;\;, \\
 \hat{P}_{K}\bullet  = \left[ p_{K}, \bullet \right]
  \;\; , \;\;
 \hat{F}_{KL} \bullet  = i \left[ \left[ p_{K}, p_{L} \right], \bullet
 \right]\;\;.
\eeqn
  In obtaining  eq.~(\ref{eq:obof}), we have set all fermionic backgrounds
 $\chi_{0}$ and  $\chi_{i}$ to zero.
As a consequence, the one-loop effective action on a generic bosonic
background is given by\footnote{ The calculation in what follows
 parallels those of refs. \cite{KEK, othercal}.}
\beqn
   W_{\one-loop} &=& 
\left( \frac{6}{2} - \frac{4}{2} -1 \right)
 Tr \log  \left( \hat{P_K} \hat{P^K} \hat{\rho}_{-}  \right)
+ \left( \frac{4}{2} - \frac{4}{2} \right)
 Tr \log  \left( \hat{P_K} \hat{P^K} \hat{\rho}_{+}  \right)
+ W_{b}+ W_{f} \;\; , \nonumber\\
\;\; \\
 W_{b} &=&  \frac{1}{4} Tr \log \left[ \left(
  \delta_{L}^{~M} + \frac{4}{(\hat{P}_{K}\hat{P}^{K})^{2}}\hat{F}_{L}^{~N}
\hat{F}_{N}^{~M} \right)  \hat{\rho}_{b \mp} \right]  \;\;,
\label{eq:W_b}  \\
 W_{f} &=&  - \frac{1}{4} Tr \log \left[  \left(1+ \frac{i}
{2\hat{P}_{K}\hat{P}^{K}}
   \Gamma^{MN} \hat{F}_{MN} \right) \hat{\rho}_{f \mp} \left(
 \frac{1+ \Gamma_{11}}{2} \right)  \right] \;\;.    \label{eq:W_f}
\eeqn
   We put  the matrix $\hat{F}_{MN}$ into the following form with
 respect to  the Lorentz indices.
\beqn
  \hat{F}_{MN} =
\left(
\begin{array}{cccccccccc}
0 &-\hat{B}_1 &0 &0 &0 &0 &0 &0 &0 &0 \\
\hat{B}_1 &0 &0 &0 &0 &0 &0 &0 &0 &0 \\
0 &0 &0 &-\hat{B}_2 &0 &0 &0 &0 &0 &0 \\
0 &0 &\hat{B}_2 &0 &0  &0 &0 &0 &0 &0 \\
0 &0 &0 &0 &0 &0 &0 &-\hat{B}_3 &0 &0 \\
0 &0 &0 &0 &0 &0 &0 &0 &-\hat{B}_4 &0 \\
0 &0 &0 &0 &0 &0 &0 &0 &0 &-\hat{B}_5 \\
0 &0 &0 &0 &\hat{B}_3 &0 &0 &0 &0 &0 \\
0 &0 &0 &0 &0 &\hat{B}_4 &0 &0 &0 &0 \\
0 &0 &0 &0 &0 &0 &\hat{B}_5 &0 &0 &0
\end{array}
\right)
\eeqn
 When the classical configuration is BPS saturated,  $\hat{F}_{MN}=0$ and
 $W_{\one-loop}$ vanishes.

\subsection{ D-string solution}

  Let us construct a few particular classical bosonic solutions of the model.
 We set  the fields lying in fundamental representation of $USp(2k)$ to zero.
Equation of motion is
\beq
[p_N , [p^M , p^N] ] = 0 \;\;\;.
\label{eq;eq of motion}
\eeq
 There are three cases  of solutions representing a D-string configuration,
 depending upon which two directions   the worldsheet extends to infinity.
  When both of  the directions are the adjoint directions, say
 $v_0$ and $v_1$, the nonvanishing components are
\beqn
 p_0 = \left( \frac{1+ \s^3}{2} \right) \otimes {\bf x}
        + \left( \frac{1- \s^3}{2} \right) \otimes (-{\bf x}^t)
\; , \;\; p_1 = \left( \frac{1+ \s^3}{2} \right) \otimes \mbox{\boldmath $\pi$}
        + \left( \frac{1- \s^3}{2} \right) \otimes (-{\bfpi}^t )\; .
\eeqn
  When  both are in the antisymmetric directions, say $v_5$ and $v_8$,
the nonvanishing
  components are
\beqn
 p_5 = \left( \frac{1+ \s^3}{2} \right) \otimes {\bf x}
        + \left( \frac{1- \s^3}{2} \right) \otimes {\bf x}^t \; ,\;\;
 p_8 = \left( \frac{1+ \s^3}{2} \right) \otimes {\bfpi}
        + \left( \frac{1- \s^3}{2} \right) \otimes {\bfpi}^t \; .
\eeqn
 When one is  in the adjoint direction, say $v_{0}$,
 and the other is in the antisymmetric
  direction,  say $v_8$,
\beqn
 p_0 = \left( \frac{1+ \s^3}{2} \right) \otimes {\bf x}
        + \left( \frac{1- \s^3}{2} \right) \otimes (-{\bf x}^t)
\;\; , \; p_8 = \left( \frac{1+ \s^3}{2} \right) \otimes {\bfpi}
        + \left( \frac{1- \s^3}{2} \right) \otimes {\bfpi}^t \;\; .
\eeqn
In above expressions, ${\bf x}$ and ${\bfpi}$ are infinite size matrices
 with the commutator $[{\bfpi}, {\bf x}]=-i$.

Let us now turn to the solutions representing two parallel D-strings and
 two anti-parallel D-strings.
 We will illustrate this by
 the most interesting case  that  the two $D$ strings are extended in
  the two directions  (  $v_5$ and $v_8$) of antisymmetric representations
  separated  by $d$ in the $v_4$ direction which is the  adjoint direction.
  The nonvanishing components are
\beqn
& & p_5 = \left( \frac{1+ \s^3}{2} \right) \otimes
        \left(
        \begin{array}{cc}
        {\bf x} & 0 \\
        0 & {\bf x}
        \end{array}
        \right)
        + \left( \frac{1- \s^3}{2} \right) \otimes
        \left(
        \begin{array}{cc}
        {\bf x}^t & 0 \\
        0 & {\bf x}^t
        \end{array}
        \right)
\nonumber\\
& & p_8 = \left( \frac{1+ \s^3}{2} \right) \otimes
        \left(
        \begin{array}{cc}
        {\bfpi} & 0 \\
        0 & {\bfpi}
        \end{array}
        \right)
        + \left( \frac{1- \s^3}{2} \right) \otimes
        \left(
        \begin{array}{cc}
        {\bfpi}^t & 0 \\
        0 & {\bfpi}^t
        \end{array}
        \right)
\nonumber\\
& & p_4 = \left( \frac{1+ \s^3}{2} \right) \otimes
        \left(
        \begin{array}{cc}
        -d/2 & 0 \\
        0 & d/2
        \end{array}
        \right)
        + \left( \frac{1- \s^3}{2} \right) \otimes
        \left(
        \begin{array}{cc}
        d/2 & 0 \\
        0 & -d/2
        \end{array}
        \right)
\;\; ,
\eeqn
for two parallel D-strings, and
\beqn
& & p_5 = \left( \frac{1+ \s^3}{2} \right) \otimes
        \left(
        \begin{array}{cc}
        {\bf x} & 0 \\
        0 & {\bf x}
        \end{array}
        \right)
        + \left( \frac{1- \s^3}{2} \right) \otimes
        \left(
        \begin{array}{cc}
        {\bf x}^t & 0 \\
        0 & {\bf x}^t
        \end{array}
        \right)
\nonumber\\
& & p_8 = \left( \frac{1+ \s^3}{2} \right) \otimes
        \left(
        \begin{array}{cc}
        {\bfpi} & 0 \\
        0 & - {\bfpi}
        \end{array}
        \right)
        + \left( \frac{1- \s^3}{2} \right) \otimes
        \left(
        \begin{array}{cc}
        {\bfpi}^t & 0 \\
        0 & -{\bfpi}^t
        \end{array}
        \right)
\nonumber\\
& & p_4 = \left( \frac{1+ \s^3}{2} \right) \otimes
        \left(
        \begin{array}{cc}
        -d/2 & 0 \\
        0 & d/2
        \end{array}
        \right)
        + \left( \frac{1- \s^3}{2} \right) \otimes
        \left(
        \begin{array}{cc}
        d/2 & 0 \\
        0 & -d/2
        \end{array}
        \right)
\;\; , \label{eq:a-para Dstrings sol}
\eeqn
for two anti-parallel D-strings.

\subsection{ force between antiparallel D-string}

  We would like to determine the scale of our spacetime given by the model.
 This can be done by computing
  the force mediating  two classical objects  which are by themselves
 a non-BPS configuration.
     We will evaluate  the $W_{b}$ and  the $W_{f}$ in the case of  the
  two antiparallel D-strings separated by distance $d$,
which have been constructed  in the last subsection.
 We  compute the force exerting with each other. 
From eq. (\ref{eq:a-para Dstrings sol}) we have
$ \hat{P}^0=\hat{P}^1=\hat{P}^2=\hat{P}^3
=\hat{P}^6=\hat{P}^7=\hat{P}^9= 0 \; $ ,
$\;\; \hat{B}_{1} =  \hat{B}_{2}=  \hat{B}_{3}=  \hat{B}_{5}  =0 \;$ ,
$\;\;  \hat{P}_{K} \hat{P}^{K} =  (\hat{P}^{4})^{2}
  +(\hat{P}^{5})^{2}  + (\hat{P}^{8})^{2}  \;$  , $\;\;
\hat{P}^{4} = \frac{d}{2} \hat{B}_{4}\;\; $
  and, after some algebra,
we obtain
\beqn
 & & \left[ \hat{P}^{5}, \hat{P}^{8} \right] = -i  \hat{B}_{4} \;\;,
\;\;  \left[ \hat{P}^{4}, \hat{P}^{5} \right] = 0 \;\;,
\;\;  \left[ \hat{P}^{4}, \hat{P}^{8} \right] = 0 \;
\;.
\eeqn

When we take trace with Lorentz indices in (\ref{eq:W_b})
and with spinor indices in (\ref{eq:W_f}), we arrive at the
following expressions:
\beqn
 W_{b} &=&  \frac{1}{2} Tr  \left[ \log  \left(
  1 - \frac{4 \hat{B}_4 \hat{B}_4 }{(\hat{P}_{K}\hat{P}^{K})^{2}}
 \right)  \hat{\rho}_{ +} \right]
\\
 W_{f} &=&  - Tr \left[ \log  \left(1 -
\frac{1}{ (\hat{P}_{K}\hat{P}^{K})^2 }
   \hat{B}_4 \hat{B}_4 \right) \hat{\rho}_{-}
+
\log  \left(1 -
\frac{1}{ (\hat{P}_{K}\hat{P}^{K})^2 }
   \hat{B}_4 \hat{B}_4 \right) \hat{\rho}_{ +}
\right]
\eeqn
   In appendix A,  the eigenvalues  of
  $\hat{B}_{4} \hat{B}_{4}$, their degeneracies and the eigenmatrices are
 determined.
  We compile the results at table $1$ for the antisymmetric eigenmatrices
and at table $2$ for the adjoint eigenmatrices.\\

\begin{tabular}[t]{|c|c|} \hline
the eigenvalue of $\hat{B}_4 \hat{B}_4$  & the degeneracy \\ \hline \hline
$ 4 $ & ${\displaystyle  k^2 - k} $ \\ \hline
$ 0 $ & ${\displaystyle  k^2 } $ \\ \hline
\end{tabular}
\begin{tabular}[t]{|c|c|} \hline
the eigenvalue of $\hat{P}_K \hat{P}^K$  & the degeneracy \\ \hline \hline
$ d^2 + 4n +2 $ & ${\displaystyle  k} $ \\ \hline
\end{tabular}

table $1$\\

\begin{tabular}[t]{|c|c|} \hline
the eigenvalue of $\hat{B}_4 \hat{B}_4$  & the degeneracy \\ \hline \hline
$ 4 $ & ${\displaystyle  k^2 + k } $ \\ \hline
$ 0 $ & ${\displaystyle  k^2 } $ \\ \hline
\end{tabular}
\begin{tabular}[t]{|c|c|} \hline
the eigenvalue of $\hat{P}_K \hat{P}^K$  & the degeneracy \\ \hline \hline
$ d^2 + 4n +2 $ & ${\displaystyle  k} $ \\ \hline
\end{tabular}

table $2$\\

\noindent
Using these tables, we obtain 
\beqn
 W_{b} &=&
\frac{k}{2}  \sum_{n=0}^{\infty} \log
 \left(  1 - \frac{16 }{( d^2 +4n +2)^{2}}
\right) \;\;,
\\
 W_{f} &=&
- 2k \sum_{n=0}^{\infty} \log
 \left(  1 - \frac{4 }{( d^2 +4n +2)^{2}}
\right)
 \;\;.
\eeqn

   Putting  all  these together, we find
\beqn
  W_{\one-loop}
&=&
 - \frac{k}{2} \log
\left[
\left( \frac{d^2}{4} \right)^{-4}
\frac{d^2/4 +1/2}{d^2/4-1/2} \left( \frac{\Gamma \left(\frac{d^2}{4} +
\frac{1}{2} \right)}{\Gamma \left(\frac{d^2}{4} \right) } \right)^8
\right]
\nonumber\\
&=&
 - \frac{k}{2}
\left\{
\frac{8}{d^6} + {\cal O} \left( d^{-8} \right)
\right\}
\;\;\; .
\eeqn
This potential provides the asymptotic behavior of the force mediating two
 antiparallel D-strings.
From this we conclude that
the dimension of spacetime is ten at least in this naive large $k$ limit.


\section{Construction of D3-brane solutions}

It is not difficult to extend the construction of the D-string solutions
in the previous section to general D$p$-brane solutions.
We will illustrate this by a D3-brane,
 two parallel D3-branes
and multiple D3-branes which are parallel to one another.

Let us first consider a D3-brane solution.
When the worldvolume extends
 in $v_5$, $v_8$, $v_6$ and $v_9$ directions,
 the nonvanishing components  are given by
\beqn
& & p_5 = \left( \frac{1+ \s^3}{2} \right) \otimes {\bf x_1}
 + \left( \frac{1- \s^3}{2} \right) \otimes {\bf x_1^t}
\;\; ,
\nonumber\\
& & p_8 = \left( \frac{1+ \s^3}{2} \right) \otimes \bfpi_1
 + \left( \frac{1- \s^3}{2} \right) \otimes \bfpi_1^t
\;\; ,
\nonumber\\
& & p_6 = \left( \frac{1+ \s^3}{2} \right) \otimes {\bf x_2}
 + \left( \frac{1- \s^3}{2} \right) \otimes {\bf x_2^t}
\;\; ,
\nonumber\\
& & p_9 = \left( \frac{1+ \s^3}{2} \right) \otimes \bfpi_2
 + \left( \frac{1- \s^3}{2} \right) \otimes \bfpi_2^t
\;\;  .
\eeqn
It is straightforward to check that this configuration satisfies
 equation of motion.
In the above expression, ${\bf x_1}$, ${\bf x_2}$, $\bfpi_1$
and $\bfpi_2$
 are operators (infinite matrices)
 with the commutators
\beq
[\bfpi_1,{\bf x_1}]=-i \sqrt{ \frac{V_4}{k} } \;\;\;, \;\;
[\bfpi_2,{\bf x_2}]=-i \sqrt{ \frac{V_4}{k} } \;\;\;.
\eeq
Here we must take the limit of $k \rightarrow \infty$ with
$V_4 / k$ fixed to $(\alpha')^2$.

Now let us calculate the value of the action.
We have
\beq
[p^5,p^8] = \s^3 \otimes i \alpha' 1_k \;\;\; ,
[p^6,p^9] = \s^3 \otimes i \alpha' 1_k \;\;\; .
\eeq
When we substitute these into the action,
\beqn
S &=& \frac{1}{g^2 (\alpha')^2} Tr
\left(
\frac{1}{2} [p^5,p^8] [p_5,p_8]
+ \frac{1}{2} [p^6,p^9] [p_6,p_9]
\right)
\nonumber\\
&\sim& \frac{1}{g^2 (\alpha')^2} V_4 = T_{\mbox{{\tiny 3-brane}}} V_4 \;\; .
\eeqn
Here $g^2$ is
regarded as string coupling $g_{\mbox{\scriptsize st}}$.
This is consistent with the D-brane action which is
given by the tension times the volume of the D-brane.
Therefore it is appropriate to think of
 the above solution as a D3-brane solution.

Next, take two parallel D3-branes 
which are separated by distance $d$ in the $v_4$ direction.
The nonvanishing components are
\beqn
& & p_5 = \left( \frac{1+ \s^3}{2} \right) \otimes
 \left(
 \begin{array}{cc}
 {\bf x_1} & 0 \\
 0 & {\bf x_1}
 \end{array}
 \right)
 + \left( \frac{1- \s^3}{2} \right) \otimes
 \left(
 \begin{array}{cc}
 {\bf x_1^t} & 0 \\
 0 & {\bf x_1^t}
 \end{array}
 \right)
\nonumber\\
& & p_8 = \left( \frac{1+ \s^3}{2} \right) \otimes
 \left(
 \begin{array}{cc}
 \bfpi_1 & 0 \\
 0 & \bfpi_1
 \end{array}
 \right)
 + \left( \frac{1- \s^3}{2} \right) \otimes
 \left(
 \begin{array}{cc}
 \bfpi^t_1 & 0 \\
 0 & \bfpi^t_1
 \end{array}
 \right)
\nonumber\\
& & p_6 = \left( \frac{1+ \s^3}{2} \right) \otimes
 \left(
 \begin{array}{cc}
 {\bf x_2} & 0 \\
 0 & {\bf x_2}
 \end{array}
 \right)
 + \left( \frac{1- \s^3}{2} \right) \otimes
 \left(
 \begin{array}{cc}
 {\bf x_2^t} & 0 \\
 0 & {\bf x_2^t}
 \end{array}
 \right)
\nonumber\\
& & p_9 = \left( \frac{1+ \s^3}{2} \right) \otimes
 \left(
 \begin{array}{cc}
 \bfpi_2 & 0 \\
 0 & \bfpi_2
 \end{array}
 \right)
 + \left( \frac{1- \s^3}{2} \right) \otimes
 \left(
 \begin{array}{cc}
 \bfpi^t_2 & 0 \\
 0 & \bfpi^t_2
 \end{array}
 \right)
\nonumber\\
& & p_4 = \left( \frac{1+ \s^3}{2} \right) \otimes
 \left(
 \begin{array}{cc}
 -d/2 & 0 \\
 0 & d/2
 \end{array}
 \right)
 + \left( \frac{1- \s^3}{2} \right) \otimes
 \left(
 \begin{array}{cc}
 d/2 & 0 \\
 0 & -d/2
 \end{array}
 \right)
\;\; .
\eeqn


Finally let us consider $N$ parallel D3-branes
which are separated in the $v_4$ and $v_7$ directions.
We denote the position of the $i$-th D3-brane
by $v_4 = d_4^{(i)}$ and $v_7 = d_7^{(i)}$.
The worldvolume extends
 in the $v_5$, $v_8$ $v_6$ and $v_9$ directions.
The nonvanishing components are 
\beqn
& & p_5 = \left( \frac{1+ \s^3}{2} \right) \otimes
 \left(
 \begin{array}{ccc}
 {\bf x_1} & & \\
           & \ddots & \\
           &        & {\bf x_1}
 \end{array}
 \right)
 + \left( \frac{1- \s^3}{2} \right) \otimes
 \left(
 \begin{array}{ccc}
 {\bf x_1^t} &  & \\
           & \ddots & \\
  & & {\bf x_1^t}
 \end{array}
 \right)
\nonumber\\
& & p_8 = \left( \frac{1+ \s^3}{2} \right) \otimes
 \left(
 \begin{array}{ccc}
 \bfpi_1 & & \\
           & \ddots & \\
  & & \bfpi_1
 \end{array}
 \right)
 + \left( \frac{1- \s^3}{2} \right) \otimes
 \left(
 \begin{array}{ccc}
 \bfpi^t_1 & & \\
           & \ddots & \\
  & & \bfpi^t_1
 \end{array}
 \right)
\nonumber\\
& & p_6 = \left( \frac{1+ \s^3}{2} \right) \otimes
 \left(
 \begin{array}{ccc}
 {\bf x_2} & & \\
           & \ddots & \\
  & & {\bf x_2}
 \end{array}
 \right)
 + \left( \frac{1- \s^3}{2} \right) \otimes
 \left(
 \begin{array}{ccc}
 {\bf x_2^t} & & \\
           & \ddots & \\
  & & {\bf x_2^t}
 \end{array}
 \right)
\nonumber\\
& & p_9 = \left( \frac{1+ \s^3}{2} \right) \otimes
 \left(
 \begin{array}{ccc}
 \bfpi_2 &  & \\
           & \ddots & \\
  & & \bfpi_2
 \end{array}
 \right)
 + \left( \frac{1- \s^3}{2} \right) \otimes
 \left(
 \begin{array}{ccc}
 \bfpi^t_2 &  & \\
           & \ddots & \\
  & & \bfpi^t_2
 \end{array}
 \right)
\nonumber\\
& & p_4 = \left( \frac{1+ \s^3}{2} \right) \otimes
 \left(
 \begin{array}{ccc}
 d_4^{(1)} & &  \\
           & \ddots & \\
  &  & d_4^{(N)}
 \end{array}
 \right)
 + \left( \frac{1- \s^3}{2} \right) \otimes
 \left(
 \begin{array}{ccc}
 - d_4^{(1)} & &  \\
          & \ddots & \\
  &  &- d_4^{(N)}
 \end{array}
 \right)
\nonumber\\
& & p_7 = \left( \frac{1+ \s^3}{2} \right) \otimes
 \left(
 \begin{array}{ccc}
 d_7^{(1)} & &  \\
           & \ddots & \\
  &  & d_7^{(N)}
 \end{array}
 \right)
 + \left( \frac{1- \s^3}{2} \right) \otimes
 \left(
 \begin{array}{ccc}
 - d_7^{(1)} & &  \\
          & \ddots & \\
  &  &- d_7^{(N)}
 \end{array}
 \right)
\;\; .
\label{eq:multi D3 sol}
\eeqn

\section{ $F$ theory on an elliptic fibered $K3$ }

We will now show that the model is able to describe the $F$ theory
  compactification on an elliptic fibered $K3$ \cite{V, Sen}. 
Our objective here is to demonstrate that the matrix model in fact derives
one of the very few exact results in critical string theory. While the
original construction of Vafa is purely geometrical in nature, our model
provides an action principle and path integrals to the $F$ theory
 compactification.
 
  In sections $4$, $5$ and $6$, 
 we have seen that our model is the matrix model of type
  $IIB$ superstrings on a large $T^{6}/{\cal Z}^{2}$ orientifold.
The coupling constant has no spacetime dependence and is a 
{\it bona fide} parameter.
  One can make the coupling space-dependent by taking the matrix $T$ dual
  in various ways to go to higher dimensional worldvolume gauge theories
 as we have already discussed in the previous sections. 
 The coupling constant then starts running with the coordinates labelling the
 quantum moduli space, {\it i.e.}  vev, which is denoted by $\vec{u}$.
  This is  in accordance with the marginal scalar
 deformation of the original action  to a type of nonlinear $\sigma$ model.
  The background field appearing through this procedure is a massless
 axion-dilaton field. The running coupling constant is, therefore,
 identified as the space-dependent  axion-dilaton background field
 $\lambda (\vec{u})$.
 
 Let $\vec{u}$ be the complex coordinates on a complex $n$-dimensional base
 space $B_{n}$. $F$ theory compactification of an elliptically fibered $C$-$Y$
 $(n+1)$ fold $M_{n+1}$ on the base $B_{n}$ is defined   by saying that
 the $u$- dependent axion-dilaton background field of type $IIB$
 superstrings on $B_{n} \times R^{9-2n.1}$ is the modular parameter of the 
 fiber $T^{2}$ as a function of $\vec{u}$.  We would like to show that this
 is in fact the case in our matrix model.
 To provide $F$ theory set-up as a reduced model for the case $n=1$,
 we are going to send the period $R$ of the four out of the six adjoint
 directions $v_{0}, \;v_{1}, \;v_{2}, \; v_{3}$ to zero and to take the matrix
 T dual. The resulting model in the limit of vanishing mass parameters
 is type $IIB$ on a large $T^{2}/{\cal Z}^{2}$ orientifold, namely on $CP^{1}$,
  equipped with sixteen $D7$-branes.
  Coupling starts running as we turn on the mass parameters.
  Following Sen\cite{Sen}, we would now like to take the scaling limit 
\beqn
  \tilde{R} &\rightarrow&  \infty\;,  \nonumber \\
  m_{i}\tilde{R} &\rightarrow&  {\rm finite} \;\; i = 1, \sim 4 \nonumber \\
   m_{i}\tilde{R} &\rightarrow&   \infty \;\; i = 5, \sim 16    \;\;.
\eeqn
  simultaneously taking the matrix $T$ dual. The second and the third lines
 of this equation come from the consistency with the $RR$ charge counting.  
 The resulting worldvolume theory around one of the four $O7$s  is
 the $d=4, {\cal N}=2$ supersymmetric $USp(2k)$ gauge theory with one
 massless antisymmetric hypermultiplet and four fundamental hypermultiplets
 with masses $m_{i}$.
   The special properties of this theory valid for all $k$ are that
 it is UV finite  and that 
at least low energy physics is the same  for all $k$
\cite{DLSASYT}. One can, therefore, deduce  the $u$ dependence of the
 coupling of  the model in the large $k$ limit  by simply looking at
 the $k=2$ case, namely,
 the $SU(2)$ susy gauge theory
 with four flavours. The $u$ dependence of the coupling
 $\lambda$ is supplied by the work of Seiberg-Witten \cite{SW}.  
 The work of Sen \cite{Sen}  shows that  the way the modular
 parameter of the bare torus  in the massless limit is dressed by the four
 mass parameters in the SW curve of the massive four flavour case is
 mathematically identical to the description of $F$ theory in
 the neighborhood of the constant  coupling.  One can therefore safely 
 conclude that the coupling $\lambda(u)$
  of the model is in fact the modular
 parameter of the spectral torus.  This is what we wanted to show.

\section{Acknowledgements}
  We thank Shinji Hirano, Toshihiro Matsuo and Asato Tsuchiya 
for helpful discussion on this subject.

\newpage
\appendix
\section{A}
In appendix A we will determine the eigenvalues
of the operators $\hat{B}_4 \hat{B}_4$ and $\hat{P}_K \hat{P}_K$.
We consider the both cases that the eigenmatrices are 
 in the adjoint and  the antisymmetric representations in $USp(2k)$.
These eigenvalues  and their degeneracy
are needed in order to calculate the one-loop effective action.

Suppose that an operator $\hat{ O}$ has an adjoint action
 on a $2k \times 2k$ matrix $ a$ :
\beq
\hat{O}\; a = [ o ,  a ] \;\; .
\eeq
Here $o$ is the $2k \times 2k$ matrix.
Let us first consider the case that
the matrix $a$ is given by eq. (\ref{eq:asym hermite matrix}).
Note that 
 the operator $\hat{B}_4 =i [\hat{P}_5 , \hat{P}_8 ]$
 is represented by the
matrix $b_4= - \s^3 \otimes \s^3 \otimes {\bf 1_{(k/2)}}$.

It is not difficult to see that the eigenvalues of $\hat{B}_4 \hat{B}_4$
are either $0$  or $4$.
For the $0$ eigenvalue we simply solve $\hat{B}_4 \; a_{(asym)}^{(0)} = 0$
and the eigenmatrices are
\beqn
& &\left( \frac{1+ \s^3}{2} \right) \otimes {\bf 1_{(2)}} \otimes H_0
 + \left( \frac{1- \s^3}{2} \right) \otimes ({\bf 1_{(2)}} \otimes H_0)^t
\;\; ,
\nonumber\\
& &\left( \frac{1+ \s^3}{2} \right) \otimes \s^3 \otimes H_3
 + \left( \frac{1- \s^3}{2} \right) \otimes (\s^3 \otimes H_3)^t
\;\; ,
\nonumber\\
& &  \s^{+} \otimes \s^1 \otimes A_1
 + \s^{-} \otimes \{ -(\s^1 \otimes A_1)^{*} \}
\;\; ,
\nonumber\\
& &  \s^{+} \otimes \s^2 \otimes A_2
 + \s^{-} \otimes \{ -(\s^2 \otimes A_2)^{*} \} \;\; .
\eeqn
Since the $(k/2) \times (k/2)$ matrices satisfy
 $H_{0 , 3}^{\dagger}=H_{0, 3}$ ,  $A_1^t=-A_1$ and $A_2^t= A_2$  ,
 the degeneracy is $k^2$.
As for the eigenvalue $4$, the solution is
\beqn
& & \left( \frac{1+ \s^3}{2} \right) \otimes
( \s^1 \otimes H_1 + \s^2 \otimes H_1 )
 + \left( \frac{1- \s^3}{2} \right)
 \otimes ( \s^1 \otimes H_1 + \s^2 \otimes H_1  )^t
\;\; ,
\nonumber\\
& & \left( \frac{1+ \s^3}{2} \right) \otimes
(  \s^1 \otimes H_2 -  \s^2 \otimes H_2 )
 + \left( \frac{1- \s^3}{2} \right) 
\otimes ( \s^1 \otimes H_2 - \s^2 \otimes H_2 )^t
\;\; ,
\nonumber\\
& &  \s^{+} \otimes ({\bf 1_{(2)}} \otimes A_0 + \s^3 \otimes A_0 )
 + \s^{-} \otimes 
\left( -({\bf 1_{(2)}} \otimes A_0 + \s^3 \otimes A_0)^{*} \right)
\;\; ,
\nonumber\\
& &  \s^{+} \otimes ( {\bf 1_{(2)}} \otimes A_3 - \s^3 \otimes A_3)
 + \s^{-} \otimes
 \left( -( {\bf 1_{(2)}} \otimes A_3 - \s^3 \otimes A_3)^{*} \right)
\;\; ,
\eeqn
and the degeneracy is $k^2-k$ 
because of $H_{1 , 2}^{\dagger}=H_{1, 2}$ and $A_{0,3}^t=-A_{0, 3}$.

Let us now calculate the eigenvalues of
the operator $\hat{P}_K \hat{P}^K = \frac{1}{4} \hat{B}_4 \hat{B}^4
+ \hat{P}_5 \hat{P}^5 + \hat{P}_8 \hat{P}^8$ .
Clearly $\hat{B}_4 \hat{B}_4$ and
$\hat{P}_5 \hat{P}^5 + \hat{P}_8 \hat{P}^8$ are simultaneously diagonalized.
When
 $\hat{P}_5 \hat{P}^5 + \hat{P}_8 \hat{P}^8$ acts on the eigenstates with
eigenvalue $4$ of $\hat{B}_4 \hat{B}_4$,
 we replace $\hat{B}_4 \hat{B}_4$ by its eigenvalue.
Let $\hat{P} \equiv \hat{P}_5 \hat{B}_4 /2 \sqrt{2}$ and
$\hat{Q} \equiv \hat{P}_8 / \sqrt{2}$.
 We obtain
 $$ [\hat{P} , \hat{Q}] = -i \;\; .$$
The eigenvalues of $\hat{P}_5 \hat{P}^5 + \hat{P}_8 \hat{P}^8
= 2(\hat{P} \hat{P} + \hat{Q} \hat{Q})$
are those of the  harmonic oscillator  and are given by 
 $ 4n+2 $ with integer $n$.
The degeneracy is $k$ for large $k$.
We summarize the results in table $1$.

\begin{tabular}[t]{|c|c|} \hline
the eigenvalue of $\hat{B}_4 \hat{B}_4$  & the degeneracy \\ \hline \hline
$ 4 $ & ${\displaystyle  k^2 - k} $ \\ \hline
$ 0 $ & ${\displaystyle  k^2 } $ \\ \hline
\end{tabular}
\begin{tabular}[t]{|c|c|} \hline
the eigenvalue of $\hat{P}_K \hat{P}^K$  & the degeneracy \\ \hline \hline
$ d^2 + 4n +2 $ & ${\displaystyle  k} $ \\ \hline
\end{tabular}

table $1$\\

\noindent
Our calculation of the effective action does not require
the case in which the eigenvalue of $\hat{B}_4 \hat{B}_4$
is zero.

Similarly, the eigenmatrices lying in the adjoint 
representation (eq. (\ref{eq:adjoint matrix}))
can be determined.
The difference is the off-diagonal degrees of freedom, which change
the degeneracy of $\hat{B}_4 \hat{B}_4$ eigenvalues.
The degeneracy of
 the $\hat{P}_K \hat{P}^K$ eigenvalues is the same in the previous case.
Summing up the adjoint case, we obtain table $2$.

\begin{tabular}[t]{|c|c|} \hline
the eigenvalue of $\hat{B}_4 \hat{B}_4$  & the degeneracy \\ \hline \hline
$ 4 $ & ${\displaystyle  k^2 + k } $ \\ \hline
$ 0 $ & ${\displaystyle  k^2 } $ \\ \hline
\end{tabular}
\begin{tabular}[t]{|c|c|} \hline
the eigenvalue of $\hat{P}_K \hat{P}^K$  & the degeneracy \\ \hline \hline
$ d^2 + 4n +2 $ & ${\displaystyle  k} $ \\ \hline
\end{tabular}

table $2$\\




\newpage

\begin{thebibliography}{99}

\bibitem{IT}
H. Itoyama and A. Tokura,
preprint hep-th/9708123, {\sl Prog. Theor. Phys.} {\bf 99} (1998) to appear.

\bibitem{Witten}
 E. Witten, {\sl Nucl. Phys.} {\bf B460} (1996)335.

\bibitem{BFSS}
T. Banks, W. Fischler, S.H. Shenker, L. Susskind, 
{\sl Phys. Rev.} {\bf D55} (1997)5112.

\bibitem{M}
 E. Witten, {\sl Nucl. Phys.} {\bf B443} (1995)85.

\bibitem{KEK}
N. Ishibashi, H. Kawai, Y. Kitazawa and A. Tsuchiya,
 {\sl Nucl. Phys.} {\bf B498} (1997)467; 
M. Fukuma, H. Kawai, Y. Kitazawa and A. Tsuchiya, preprint hep-th/9705128.

\bibitem{Nambu}
Y. Nambu,
 p. 1 in ``Quark Confinement and Field 
Theory" (John Wileys \& Sons, New York, 1977).

\bibitem{Bars}
I. Bars,
{\sl Phys. Lett.} {\bf 245B} (1990) 35.

\bibitem{FFZ}
D.B. Fairlie, P. Fletcher and C.Z. Zachos,
{\sl J. Math Phys.} {\bf 31} (1990) 1088.

\bibitem{Schild}
A. Schild,
{\sl Phys. Rev.} {\bf D16} (1977)1722.

\bibitem{subdev}
V. Periwal, {\sl Phys. Rev.} {\bf D55} (1997) 1711; 
L. Motl, preprint hep-th/9612198; 
N. Kim and S.J. Rey, {\sl Nucl.Phys.} {\bf B504} (1997) 189; 
R. Dijkgraaf, E. Verlinde and H. Verlinde, 
{\sl Nucl.Phys.} {\bf B500} (1997) 43; 
A. Fayyazuddin and D.J. Smith, preprint hep-th/9703208; 
T. Banks and L. Motl, preprint hep-th/9703218; 
N. Kim and S.J. Rey, preprint hep-th/9705132; 
S. Sethi and L. Susskind, preprint hep-th/9702101; 
T. Banks and N. Seiberg, preprint hep-th/9702187; 
D. A. Lowe, {\sl Phys .Lett. } {\bf 403B} (1997) 243
; 
S.J. Rey, {\sl Nucl.Phys.} {\bf B502} (1997) 170
; 
P. Horava, {\sl Nucl.Phys.} {\bf B505} (1997) 84; 
%
%
M. Li, {\sl Nucl. Phys.} {\bf B499} (1997) 149; 
M.B. Green and M. Gutperle, {\sl Phys .Lett. } {\bf B398} (1997) 69; 
I. Chepelev, Y. Makeenko and K. Zarembo,
{\sl Phys. Lett.} {\bf 400B} (1997)43; 
A. Fayyazuddin and D.J. Smith, 
{\sl Int. J. Mod. Phys. Lett.} {\bf A12} (1997) 1447; 
A. Fayyazuddin, Y. Makeenko, P. Olsen, D. J. Smith and K. Zarembo,
preprint hep-th/9703038; 
T. Yoneya, {\sl Prog. Theor. Phys.} {\bf 97} (1997) 949; 
B. Sathiapalan, {\sl Int. J. Mod. Phys. Lett.} {\bf A12} (1997) 1301; 
C.F. Kristjansen and P. Olesen, {\sl Phys. Lett.} {\bf B405} (1997) 45; 
L. Chekhov and K. Zarembo, {Int. J. Mod. Phys. Lett.} {\bf A12} (1997) 2331; 
I. Chepelev and A. A. Tseytlin, preprint hep-th/9705120; 
K.-J. Hamada, {\sl Phys.Rev.} {\bf D56} (1997) 7503; 
O. A. Solovev, preprint hep-th/9707043; 
N. Kitsunezaki and J. Nishimura, preprint hep-th/9707162; 
H. Sugawara,  preprint hep-th/9708029; 
S. Hirano and M. Kato, preprint hep-th/9708039; 
B. P. Mandal and S. Mukhopadhyay, preprint hep-th/9709098; 
I. Oda, preprint hep-th/9710030; 
A.K. Biswas, A.K. Kumar and G. Sengupta, preprint hep-th/9711040; 
T. Suyama and A. Tsuchiya, preprint hep-th/9711073.





\bibitem{GSano} 
M. B. Green and J. H. Schwarz, {\sl Phys. Lett.} {\bf 149B} (1984) 117.

\bibitem{GHMR}
D. J. Gross, J. A. Harvey, E. Martinec and R. Rohm,
{\sl Phys. Rev. Lett.} {\bf 54} (1985)502,
{\sl Nucl. Phys.} {\bf B256} (1985)253,
{\sl Nucl. Phys.} {\bf B267} (1986)75.

\bibitem{PW}
 J. Polchinski and E. Witten, {\sl Nucl. Phys.} {\bf B460} (1996)525.

\bibitem{V}
C. Vafa,
{\sl Nucl.Phys.}{\bf B469}(1996)403.

\bibitem{SL(2Z)}
C. M. Hull and P. K. Townsend,
 {\sl Nucl. Phys.} {\bf B438} (1995)109

\bibitem{tH}
G. 't Hooft,
{\sl Nucl.Phys.} {\bf B72} (1974) 461.

\bibitem{WT}
W. Taylor IV , {\sl Phys. Lett.} {\bf B394} (1997)283;
O.J. Ganor, S. Ramgoolam and W. Taylor IV,
{\sl Nucl.Phys.}{\bf B492}(1997)191.

\bibitem{BSS}
L. Brink, J. Scherk and J. H. Schwarz,
{\sl Nucl.Phys.} {\bf B121} (1977) 77.


\bibitem{PR}
C. N. Pope and L. Romans,  {\sl Class. Quantum Grav.}{\bf 7} (1990) 97.

\bibitem{Mtwist}
U.H. Danielsson and G. Ferretti,
{\sl Int. J. Mod. Phys.} {\bf A12} (1997)4581; 
L. Motl,
preprint hep-th/9612198; 
N. Kim and S.J. Rey,
{\sl Nucl.Phys.} {\bf B504} (1997)189.

\bibitem{Ba}
S. Kachru and E. Silverstein, {\sl Phys. Lett.} {\bf 396B} (1997)70;
D. A. Lowe, {\sl Nucl.Phys.} {\bf B501} (1977) 134,
{\sl Phys. Lett.} {\bf 403B} (1997)243;
T. Banks, N. Seiberg and E. Silverstein,
{\sl Phys. Lett.} {\bf 401B} (1997)30.


\bibitem{GS}
M. B. Green and J. H. Schwarz, 
{\sl Phys. Lett.} {\bf 151B} (1985)21 

\bibitem{IM}
 H. Itoyama and P. Moxhay, {\sl Nucl. Phys.} {\bf B293} (1987)685

\bibitem{AIKT}
 Y. Arakane, H. Itoyama, H. Kunitomo and A. Tokura, 
{\sl Nucl. Phys.} {\bf B486} (1997)149.

\bibitem{ZWZ}
B. Zumino, W. Yong-Shi and A. Zee, {\sl Nucl.Phys.} {\bf B239} (1984) 477 .

\bibitem{othercal}
I. Chepelev, Y. Makeenko and K. Zarembo,
{\sl Phys. Lett.} {\bf 400B} (1997)43 ; 
I. Chepelev and A. A. Tseytlin, preprint hep-th/9705120.


\bibitem{Sen}
A. Sen,
{\sl Nucl.Phys.}{\bf B475}(1996)562.


\bibitem{DLSASYT}
M.R. Douglas, D.A. Lowe, J.H. Schwarz, 
{\sl Phys.Lett.}{\bf B394}(1997)297; 
O. Aharony, J. Sonnenschein, S. Yankielowicz and S. Theisen, 
{\sl Nucl.Phys.} {\bf B493} (1997) 177 .

\bibitem{SW}
 N. Seiberg and E. Witten, 
{\sl Nucl. Phys.} {\bf B426} (1994)19 ,
{\sl Nucl. Phys.} {\bf B431} (1994)484.
\\


\end{thebibliography}
\end{document}